%% file: main.tex
\documentclass[12pt, draftclsnofoot, onecolumn]{IEEEtran}
\usepackage{cite}
\usepackage{setspace}
\usepackage{graphicx}
\usepackage{array}
\usepackage{tikz}
\usepackage{tikz-3dplot}
\usetikzlibrary{shapes.geometric}
\usetikzlibrary{decorations.pathreplacing}
\usetikzlibrary{positioning}
\usetikzlibrary{patterns}
\usepackage{pgfplots}
\usepgfplotslibrary{fillbetween}
\usepackage{pstool}  % Newer versions of latex, remove psfrag and auto-pst-pdf and use pstool instead.
\usepackage{mathtools}
\usepackage[linesnumbered,algoruled,boxed,lined]{algorithm2e}
\usepackage[absolute]{textpos}
\usepackage{threeparttable}
\usepackage[strict]{changepage}
\usepackage{amssymb}
\usepackage{comment}
\usepackage{bbm}
\usepackage{hhline}
\usepackage{lipsum} 
\usepackage{enumitem}
\usepackage{xcolor}

\usepackage{amsthm}

\newcommand{\review}[1]{{\color{black}#1}}
\newcommand{\hl}[1]{{\color{black}#1}} % Faris added this for Faris's changes

\ifCLASSINFOpdf
\else
\fi
\usepackage{amsmath}
\ifCLASSOPTIONcompsoc
  \usepackage[caption=false,font=normalsize,labelfont=sf,textfont=sf]{subfig}
\else
  \usepackage[caption=false,font=footnotesize]{subfig}
\fi
\pgfplotsset{compat=1.14} % arXiv supports 1.14

\begin{document}
\title{Deep Learning for Multi-User Proactive Beam Handoff: A 6G Application}
% For journal or letter
\author{Faris~B.~Mismar,~\IEEEmembership{Senior~Member,~IEEE},% <-this % stops a space
       ~Alperen~Gundogan,~\IEEEmembership{Member, IEEE},% <-this % stops a space
       ~Aliye~\"{O}zge~Kaya,~\IEEEmembership{Senior~Member,~IEEE},% <-this % stops a space
       ~and Oleg~Chistyakov,~\IEEEmembership{Member,~IEEE}% <-this % stops a space
%}% <-this % stops a space
\thanks{The authors are with Nokia Bell Labs Consulting and Nokia Bell Labs. Email: faris.mismar@bell-labs-consulting.com and \{alperen.gundogan, ozge.kaya, oleg.chistyakov\}@nokia-bell-labs.com.}% <-this % stops a space
%\thanks{This paper is an expanded version of \cite{paperarXiv}.}
}%

\maketitle
\begin{abstract}
This paper demonstrates the use of deep learning and time series data generated from user equipment (UE) beam measurements and positions collected by the base station (BS) to enable handoffs between beams that belong to the same or different BSs. We propose the use of long short-term memory (LSTM) recurrent neural networks with three different approaches and vary the number of lookbacks of the beam measurements to study the performance of the prediction used for the proactive beam handoff.   \review{Simulations show that while UE positions can improve the prediction performance, it is only up to a certain point.  At a sufficiently large number of lookbacks, the UE positions become irrelevant to the prediction accuracy since the LSTMs are able to learn the optimal beam based on implicitly defined positions from the time-defined trajectories.}
\end{abstract}

\begin{IEEEkeywords}
deep learning, transfer learning, predictive, beamforming, handoff, radio resource management
\end{IEEEkeywords}

% \fi
\IEEEpeerreviewmaketitle

%%%%%%%%%%%%%%%%%%%%%%%%%%%%%%%%%%%%%%%%%%%%%
\section{Introduction}\label{sec:introduction} % must be 3 columns

%\IEEEPARstart{T}{he} 

%I wish you the best of success.

 
Since the introduction of the fifth generation of wireless networks (5G), user equipment (UEs) have become associated with serving beams instead of serving base stations.  For stationary UEs, beams can improve the signal strength as measured by these UEs in contrast to a conventional BS with no beamforming. Beamforming in 5G allows the UE reference symbols to benefit from the beamforming gain---an opportunity the reference symbols missed in prior generations of wireless networks.  However, several challenges emerge with mobility. %, especially with communication at the millimeter wave (mmWave) frequency range or higher (e.g., terahertz waves). 
These challenges can be summarized in either blockage or handoff interruption (or both).  {A proactive approach to beam assignment to moving UEs can help reduce the impact of these challenges and improve the reliability in these networks \cite{8646438}.}

Handoffs are radio resource management (RRM) mobility procedures used to transfer the UE session from one base station (BS) to the other.  In 5G, the concept of beam switching (or \textit{handoff}) is introduced where a UE session {is served by a different beam}
%obtains a different serving beam 
instead of its current one based on the radio measurements of {the reported beams from UE.}
%these beams. 
Beam switching  can happen within the same BS (intra-BS) or between two different BSs (inter-BS).  While it is straightforward for the UE to switch to a different beam in an intra-BS beam switching  scenario, a random access procedure is often needed for the inter-BS beam switching scenario.  In this case, the UE detects the most performance-optimal beam and sends back a physical random access mapped to the identifier of the beam. This procedure is not instantaneous since the UE has to wait for the random access procedure to take place before it can proceed \cite{3gpp38213}.  %
% The interruption time of service due to the random access procedure may depend on the selected beam---whether the switching is intra- or inter-BS type---and the air interface layer that is carrying out the mobility procedure (i.e., upper- or lower-layer mobility). 
During this time, further radio degradation in the current serving beam may happen and the UE may indicate a radio link failure (RLF) and perform a re-establishment procedure \cite{3gpp38331}. This procedure leads to longer service interruption as the UE now has to perform a random access procedure. It becomes an opportunity for an intelligent data-driven approach to proactively prepare the beam for every UE in transition.  This is in order to avoid interruption in the flow of its data as a result of this beam switching.

\begin{figure}[!t]
%\begin{adjustwidth}{.25in}{0cm}
\centering
%\resizebox{0.35\textwidth}{!}{\input{figures/overall.tikz}}%
\resizebox{0.45\textwidth}{!}{\input{figures/overall.tikz}}%
%\end{adjustwidth}
\caption{Next-generation base station uses an edge node for storage (solid) and compute (dashed) to proactively assign beams to UEs using radio measurements.}
\label{fig:overall}
\end{figure}
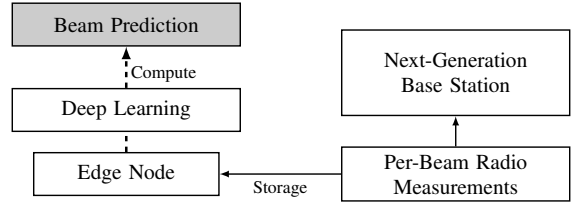

With the development of technology beyond 5G, the introduction of artificial intelligence and machine learning (AI/ML) is inevitable: the data-driven approaches to solve problems that are mathematically intractable using AI/ML have witnessed established roots to industry standards. Specifically, there are study items on enhancements for data collection for 5G \cite{rp201620} and AI/ML air interface \cite{rp213599}. Deep learning, a specialized AI/ML technique, has shown viable reusability and reduction in computation cycles through ``transfer learning'' \cite{5288526}.  Transfer learning is a deep learning approach that uses a pre-trained model for one task as a starting point for another that performs a similar task, which can also be re-trained.  It has been used with measurable success, as we also validate in this paper.  Deep learning methods are a specialized form of machine learning that uses neural network architectures.  Owed to their non-linear activation functions, these learning methods can learn arbitrarily complex relationships between inputs and outputs and have a certain level of robustness against noise.  Also, due to the multiple layers of neural networks (or \textit{depth}) and number of nodes in each layer (or \textit{width}), they can autonomously extract learning features making the task of learning more time-efficient.  Deep learning can perform time series classification.  In this type of classification, training data is time-indexed and the learning must abide to the temporal relationship between observations in the data.  %This training data has multiple learning features and are input to the deep learning model with a single output representing the class labels, and thus is a multivariate input univariate output time series classification learning problem.%

The use of deep learning in performing predictive handoff or beam switching serves other purposes with regards to the evolution towards the sixth generation of wireless networks (6G): on the one hand, it makes a solid use case for multi-access edge computing, where both data storage and compute power are brought near to the base station \cite{9500204}.  On the other hand, successful predictions can eliminate any potential disruption in the flow of data to the UE \cite{9199558, 8665922}, thus improving on the throughput and latency.  These two measures remain quintessential performance measures as wireless networks evolve towards 6G.  Fig.~\ref{fig:overall} shows how an edge compute and storage node can benefit the next-generation base stations in storing data and performing AI/ML-driven computations related to beam prediction and proactive beam handoffs.  

%\vspace*{-2em}
\subsection{Related Work}
%Prior work related to proactive mobility radio resource management procedures has been studied extensively in recent literature.

A single user moving in a trajectory of two base stations (BSs) was studied in \cite{8646438}, \cite{8736709}.  In \cite{8646438}, the
objective was to minimize blocking probability using a gated recurrent unit (GRU) deep learner.  We study multiple users moving in different trajectories with more than two BSs. Further, GRUs have the potential to overfit, a problem that can be avoided with models with more complexity, such as the long short-term memory deep neural network, which we propose.  In \cite{8736709}, the assumption on the trajectory is that it is a fixed turning scene with a predetermined finite angle and radius.  In our case, the trajectories are not predetermined and are based on two-dimensional streets with the motion of UEs being stochastic in both the direction and speed.  \hl{Further, several papers successfully used these long short-term memory deep neural networks in areas not related to wireless networks such as} \cite{a15110393}, \hl{where different data-driven approaches were applied to proactively determine the useful life of Lithium-ion batteries and thus avoid probable damages.}

The industry standards \cite{3gpp38331, 38321} for mobility procedures can be divided as higher- and lower-layer mobility. Higher-layer mobility requires the Radio Resource Configuration (RRC) protocol to initiate handover procedure. The configuration to the UE can be provided earlier (i.e., conditional handover) or on time (i.e., baseline handover and dual active protocol stack handover) \cite{3gpp38331}. The handover in RRC layer is triggered if the beam received signal power of non-serving BS is higher than that of the beam of a serving BS by a certain offset and for a given duration \cite{3gpp38331}.  In lower-layer mobility (LLM), the handover decision is performed by medium access control (MAC) layer of the air interface protocol stack based on the received beam measurements.  MAC layer uses certain signaling information known as the ``control elements,'' which are sent to the UE to trigger the
{serving beam change. UEs can be configured to report the beams of both serving BS and non-serving BS, and thus a beam change can be performed in both intra-BS or inter-BS scenarios.}
%intra-BS beam change including non-serving BSs.  
The configuration of the {LLM} 
%this
mobility procedure is provided to the UE during the RRC procedure in advance {as in conditional handover}. %The enhancements on LLM is one of the objectives of 3GPP Rel. 18 to facilitate inter-BS beam change \cite{rp221799} in order to reduce interruption time during handover. The CSI measurement configuration can be extended where the UEs can report the measurements of beams associated with the non-serving BSs in addition to the serving BSs. 
%LLM aims to reduce the interruption time by a significant order of magnitude of that of RRC mobility. The performance of LLM is compared with baseline handover, CHO and DAPS in paper \cite{nokiawp} and it is shown that LLM reduces the number of radio link problems by a factor close to a fifth of that of the baseline handover. 
In inter-BS mobility the decision to trigger LLM is made by the serving BS and conditions for this decision have not been defined in the standards. %Therefore, the conventional mobility robustness optimization (MRO) solutions for L3 mobility as proposed by authors in \cite{6125335} are not applicable for LLM. 
Thus, this provides room for novelty in terms of how this procedure can be implemented and we propose the predictive capability using prior measurements.

% it can be quite challenging for the network to define the rule based approach (e.g.  A3 event) for L1 inter-BS mobility by considering the radio characteristics in FR2 and the fluctuations on L1-RSRP values. Therefore, in this paper, we investigated to predict the next best beam index of the UE based on the previous measurements collected by the CSI measurements reports that include both serving and non-serving BS beam measurements.

% The handover procedures defined in these standards 3GPP (e.g. baseline handover or conditional handover) so far can be controlled by certain events and the conditions of those events are defined and are sent to the UE as a part of RRC procedure. For example; if the signal power of the non-serving BS is higher than the signal power of the serving BS by an offset and this condition holds for certain interval, the A3 event can be triggered to trigger the handover. Note that, the mobility robutness optimization (MRO) methods can be performed for classical L3 handover procedures in order to optimize the conditions for handover e.g. BS individual offset, Time-to-trigger, etc. [ref to Ingo's work]. 

An alternative to beam switching was proposed in \cite{9199558} through use of deep learning to perform band switching.  In essence, channels of different frequency bands transmitted off the same BS (e.g., millimeter wave (mmWave) and sub-6 GHz) possess certain correlation characteristics that can be exploited to predict whether a UE switching from one frequency band to another would be successful.  The use of UE coordinates was suggested as a means to improve the prediction performance.  While we also propose the use of UE reported coordinates, we limit our dependence on the coordinates through the use of time series since time series captures the best serving beam of a given UE in its trajectory as a function of time.  This implicitly captures the channel impact due to the location of the UE.

\hl{We note that in this paper, it is the 6G evolution of the base station to include compute capabilities that we exploit, and  not any specific radio measurement changes, though these may prove helpful once the standards are finalized.}

\subsection{Contributions}

This paper makes the following specific contributions:

\begin{enumerate}
\item Demonstrate the importance of per-UE deep learners in BSs as opposed to one single learner per BS.
\item Show that knowledge of UE coordinates can improve prediction performance {but up to a certain point in history}.
\end{enumerate}%

%%%%%%%%%%%%%%%%%%%%%%%%%%%%%%%%%%%%%%%%%%%%%
\vspace*{-0.5em}
\section{System Model}\label{sec:sysmodel}
%%%%%%%%%%%%%%%%%%%%%%%%%%%%%%%%%%%%%%%%%%%%%

We consider a system composed of a multiple base stations (BS) with multiple transmit (i.e., downlink) and a single receive (i.e., uplink) antenna.  Let there be $M_T$ transmit antennas per BS.  UEs are independently scattered in the service areas of these BSs and are in motion defined by a trajectory.    Let the set $\mathcal{B}$ be defined as the set of BSs that cover the entire service area.

The received signal of the $i$-th UE on the $b$-th beam from the serving BS for a given subcarrier is therefore given by:
\begin{equation}
    y_i = \mathbf{h}_{b,i}^\ast \mathbf{f}^\star_{i,b} x_i + \sum_{j \neq i} \mathbf{h}_{b,j}^\ast \mathbf{f}^\star_{i,b} x_j + n_i,\qquad i \in \{1,2,\ldots,u\}
    \label{eq:optimal_bf}
\end{equation}% 
\noindent where $x_i$ is the transmitted signal of the $i$-th UE normalized such that $P_{x_i} \coloneqq \mathbb{E} \vert x_i \vert ^2 = 1$ and $\mathbf{h}_{b,i}\in\mathbb{C}^M$ is the channel vector connecting the unique beam $b$ to the $i$-th UE.  {We assume that the channel state information represented by these channel vectors $\mathbf{h}_{b,\cdot}$ are perfectly known at the serving BS and UE,} which is a common practice for transmissions at the mmWave frequency range where time division duplex (TDD) mode of operation is used and channel reciprocity is exploited.  Further, we define these beams in a way that make them uniquely identifiable across all BSs.  Let there be a total of $u > 0$ UEs in the system.  The power-optimal beamforming vector $\mathbf{f}^\star_{i,b}$ is selected from a quantized analog beamforming codebook $\mathcal{F}$ with a finite cardinality:
\begin{equation}
    \mathbf{f}_{i,b}^\star = \underset {\mathbf{f}\in\mathcal{F}}{\arg\, \max}\; \vert \mathbf{h}_{b,i}^\ast \mathbf{f} \vert ^ 2.
\end{equation}% 

The last two terms in \eqref{eq:optimal_bf} correspond to the inter-beam interference and additive white noise, which is sampled from a zero-mean Gaussian probability density function and a variance of $\sigma^2$.  The received signal power can be computed as $P_{y_i} = \vert h_{b,i}^\ast f_{i,b}^\star\vert^2$, which when measured on the reference symbols on the synchronization channels per beam, is known as the ``SS-RSRP'' \cite{3gpp38300}.

% This allows us to define the average signal to interference plus noise (SINR) quantity---in the ergodic sense---as measured by the $i$-th UE as:
% \begin{equation}
%     \gamma_{i} \coloneqq \frac{\mathbb{E} \vert \mathbf{h}_b^\ast \mathbf{f}^\star_{i,b} \vert ^2}{\sum_{j \neq i} \mathbb{E} \vert \mathbf{h}_b^\ast \mathbf{f}^\star_{j,b}\vert ^2 + \sigma^2}
% \end{equation}

\textbf{Fading:} Since the UEs are in motion, they are expected to experience fading.  In the presence of fading, we assume that the channel observes two types of fading: 1) large scale: shadow fading (i.e., log-Normal) and 2) small scale: Doppler fading depending on the movement speed of the UEs.  The impact of the fading is captured in the channel state information.

\textbf{Beam switching:} As mentioned in Section~\ref{sec:introduction}, the change of the serving beam (i.e., the beam switching (or \textit{handoff}) procedure) can be implemented at several layers in the air interface protocol stack.  However, as in \eqref{eq:optimal_bf}, the power-optimal beam is computed through a search over a codebook while the channel is assumed constant (i.e., within its coherence time).  With beamforming, the channel coherence time increases due to directional reception \cite{1343892}.  The beam coherence time for a moving UE is approximately given by \cite{7742901,9199558}:
\begin{equation}\label{eq:coherence}
    t_{\text{coherence}, i} \approx \frac{D_i}{v_i \sin \alpha_i} \frac{\Theta}{2},
\end{equation}
where $D_i$ is the Euclidean distance of the $i$-th UE from the serving BS and $\Theta$ is the beamwidth of the serving beam measured in radians, $v_i$ is the speed of the UE on the trajectory, and $\alpha_i$ is the angle between the direction of travel and the direction of the BS.  UEs are expected to have different coherence times since they have different locations and distances from their respective serving BSs.

%Further, the beam handoff procedure can be performed both within the same serving BS (intra-BS beam management) or UE can connect to the beam of the non-serving BS (inter-BS beam management) as the measurements reported by the UE include the beam measurements of both serving and non-serving BSs. The beam switch is determined by the serving BS and MAC CE message is sent to the UE to perform beam switch.

%%%%%%%%%%%%%%%%%%%%%%%%%%%%%%%%%%%%%%%%%%%%%
\vspace*{-0.75em}
\section{Problem Formulation}\label{sec:problem}
%%%%%%%%%%%%%%%%%%%%%%%%%%%%%%%%%%%%%%%%%%%%%

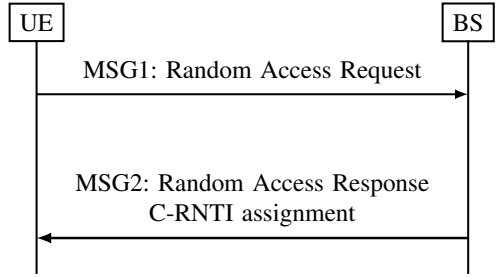
\begin{figure}[!t]
%\begin{adjustwidth}{.25in}{0cm}
\centering
\resizebox{0.4\textwidth}{!}{\input{figures/signaling_assign.tikz}}%
%\end{adjustwidth}
\caption{Assignment of the C-RNTI as part of the random access procedure defined in the standards.}
\label{fig:crnti}
\end{figure}

The industry standards \cite{38321} provide a UE-specific identifier known as the BS ``radio network temporary identifier'' (C-RNTI).  The C-RNTI is assigned to the UE during its random access procedure by the serving BS as shown in Fig.~\ref{fig:crnti}.  It remains with the UE as long as it is served by that BS.   This temporary identifier makes building a deep learning model per BS-UE pair possible.  Further, standards \cite{3gpp38300} also provide identifiers per beam per BS.  However, as there are multiple BSs, we propose one way to make the beam identifier unique across the network through a binary shift left and add operation, with the shift amount being equal to $\log_2 \vert \mathcal{B} \vert$.

This motivation allows us to formulate the problem of proactively selecting the optimal beam at the radio frame number $n$ and the $i$-th UE:%
\begin{equation}
    b_i^\star[n + 1] \coloneqq \underset{b\in\mathcal{B},\ \ell\in[0,\vert\mathcal{L}\vert - 1]}{\arg\,\max}\; \hat M_b(n; i,\ell)
    \label{eq:beam_selection}
\end{equation}%
where $\hat M_b(\cdot)$ is the deep learning estimate of the likelihood of the predicted beam to be for the optimal $b$-th beam using $\ell$ lookback beam measurements assigned to the $i$-th user.  The lookback values selected from the ordered set $\mathcal{L}$. This formulation allows us to treat both intra- and inter-BS beam handoff alike.

From \eqref{eq:beam_selection}, we are training a multi-class classifier based on time series to predict the optimal beam for $(n + 1)$, which means the beam for the \textit{next} radio frame.  While the offset term can be set to more than $1$, we choose $1$ which corresponds to the next radio frame. %
% which corresponds to the next radio frame (or $10$ ms).  
{The main reason of predicting the optimal beam for the next radio frame only is that the beam switching (of the UE) to the target beam has to be completed within $1$~ms for LLM according to the standards \cite{3gppevm}.}  %
% \faris{In perspective, the reporting interval for beam measurements can
% %\alperen{be set as lo}
% be as low as \alperen{I dont understand the message in this sentence} high as $120$ ms for RRC-layer mobility \cite{3gpp38331}}.

%Compared to the RRC-layer mobility, handover procedure of the UE after it receives the handover message from the network is assumed to be 80ms \cite{3gppevm}, therefore for RRC layer mobility, the network needs to have  longer prediction window. In addition, shorter prediction interval can increase the accuracy of the predictions [any ref?]. Therefore, proactive beam handoff mechanism proposed in the paper suits well with the industry assumptions}
%Because of this, any offset greater than $1$ may conflict with the standards settings causing the RRC-layer mobility to simultaneously trigger (i.e., a race condition).

\vspace*{-0.5em}
\section{Deep Learning for Time Series}\label{sec:time_series}
%%%%%%%%%%%%%%%%%%%%%%%%%%%%%%%%%%%%%%%%%%%%%

In this section we explain how time series data $y_t$ can be used in deep learning.  Particularly, we study the multi-variate time series case (i.e., where $y_t$ is a function of many learning features measured at time $t$ for a given UE).

\vspace*{-0.5em}
\subsection{Time Series} 

Time series $y_t$ is a sequence of data points indexed in \textit{discrete} time $t$.  This time can be in steps as small as milliseconds {depending on the network configurations}, which is the case when data points are collected from radio measurements.    The $v$-th learning feature ($v\in\{0,1,\ldots,N-1\}$) is denoted in a column vector format as: $\mathbf{x}^{(v)} \coloneqq [x_t^{(v)}]_{t = 0}^{M-1} \in \mathbb{R}^M$, where $x_t^{(v)}$ is the value of the $v$-th learning feature measured at the time instance $t$ for the UE being served during this time instance.

In the context of our problem, $y_t$ is the power-optimal beam identifier as measured at the radio frame $t$ for a UE moving along an arbitrary trajectory.  It is used to construct a column vector $\mathbf{y}\coloneqq[y_t]_{t = 0}^{M-1}\in\mathbb{R}^M$.  All $\mathbf{x}^{(v)}$ and $\mathbf{y}$ are joined on the parameter $t$ to construct the time-indexed dataset.  At any given $t$ and for a given UE, this dataset is a design matrix of joined $[\mathbf{x}^{(v)}\,\vert\, \mathbf{y}]$.  For all UEs, a similar matrix can be constructed from this by stacking these design matrix instances based on time.  This stacked matrix is referred to as $\mathbf{D}\in\mathbbm{R}^{Mu\times n}$, where $u>0$ is the number of UEs in the network as defined in \eqref{eq:optimal_bf}.
Given that the dataset contains a supervisory signal, $\mathbf{y}$, we may reconstruct this multi-variate time series problem as a supervised learning problem with the supervisory signal being a time-shifted version of $\mathbf{y}$.   We discuss this as well as the impact of feature engineering further in this section.%

\textbf{Horizon} is the number of time shifts in the \textit{future} that we would like to predict.  The horizon is a function of the learning features extending back in time, while it looks forward in time.  In \eqref{eq:beam_selection}, our horizon is set to $1$ as motivated earlier.

\textbf{Embedding} is transformation of a single dimension of temporal sequence to a multi-dimension space.  By setting up a sequence of delays, we can treat each \textit{past} value as an additional spatial dimension in the input space.  If we look back in history by $\delta > 0$ time steps, then we have an embedding dimension of $\delta$.  % https://www.cs.cmu.edu/afs/cs/academic/class/15782-f06/slides/timeseries.pdf

\vspace*{-0.5em}
\subsection{Feature Engineering}
There are three types of feature engineering that we consider: 1) shifting and lagging, 2) differencing, and 3) scaling.

\textbf{Shifting and lagging:} To improve the predictability of a time series, a time shift of the dataset $\mathbf{D}$ is applied block-wise with past and future shift values.  That is, the dataset becomes $[\mathbf{D}_{t-k}\vert\ldots\vert\mathbf{D}_t\vert\ldots\vert\mathbf{D}_{t+\ell}]$, where $k>0$ is the lag and $\ell > 0$ is the lead time shift.  This causes the dataset to have an additional total of $\ell + k$ column vectors, which represent the embedding dimension as defined earlier.   This time shift operation also creates undefined values which are often dropped from the dataset causing a reduction in the number of \textit{rows} by $\ell + k$.

\textbf{Differencing:} A stationary time series is a series the statistics of which do not depend on the time at which the series is observed.  Differencing helps reduce any trend in the time series, and therefore introduces stationarity and improves predictability.  Let us define the $j$-th order difference $d^{(j)}(t) \coloneqq x_t - x_{t - j}, j\in \{1,2,3,\ldots\,M\}$ and apply it column-wise on $\mathbf{D}$.  This leads to a new matrix $[\boldsymbol\Delta_{t-k}\vert\ldots\vert\mathbf{0}\vert\ldots\vert\boldsymbol\Delta_{t+\ell}]$, with $\ell + k + 1$ columns.  Combined with the shifting and lagging step, the dataset $\mathbf{D}$ thus has a total of $2(\ell + k + 1)$ column vectors, each of the dimension $Mu$ for a shape $\mathbf{D}\in\mathbb{R}^{(M - (\ell + k) )\times 2(\ell + k + 1)}$. We set $N\coloneqq 2(\ell + k + 1)$ and $M^\prime \coloneqq Mu - (\ell + k )$ for simplification and thus $\mathbf{D}\in\mathbb{R}^{M^\prime \times N}$ after the differencing operation.

\textbf{Scaling:} Scaling the learning features is done through the transformation function $\varphi : \mathbf{D} \mapsto \mathbf{\tilde{D}}$, where the scaling of values between $[0,1]$ is essential to avoid driving the neural network activation functions deep into saturation.  We do not scale the supervisory signal vector.%

\vspace*{-0.5em}
\subsection{Supervised Learning and Prediction}
While the dataset $\mathbf{D}$ contains a supervisory signal $\mathbf{y}$ which is the time series at present time $t$, predictions are interested in \textit{future} values of $\mathbf{y}$.  To address this discrepancy, we define a lookahead $L \in \{0,1,2,\ldots\}$ where the supervisory signal becomes $\mathbf{y}_L \coloneqq[y_{t+L}]_{t = 0}^{M^\prime-1}\in\mathbb{R}^{M^\prime}$.  This way, the supervised learning problem is established with respect to a lookahead value of the supervisory signal $\mathbf{y}_L$.%
 
Combined with feature engineering, we can construct the time-series supervised learnable dataset $\mathbf{D} \coloneqq [\mathbf{\tilde D}\,\vert\, \mathbf{y}_L]$.  We use this definition of $\mathbf{D}$ as our learnable dataset moving forward, which is later reshaped into a tensor, to train the deep learning model as we see later in this section. %

\textbf{Prediction:} If we denote the true supervisory signal at time $t$ as $\mathbf{y}_L$, then the predicted signal is denoted as $\mathbf{\hat y}_L$.  
Since the supervisory signal $\mathbf{y}_L$ is categorical, the prediction problem becomes a minimization of the categorical cross-entropy of the probability density $p(\cdot)$ of $\mathbf{\hat y}_L$, which is defined as:
\begin{equation}\label{eq:loss}
    L(\mathbf{y}_L, \mathbf{\hat y}_L) \coloneqq -\sum_{i = 0}^{M^\prime-1} \sum_{b\in\mathcal{F}} \mathbbm{1}[[\mathbf{y}_L]_i = b] \cdot \log_2 p([\mathbf{\hat y}_L]_i = b).
\end{equation}
 
% The difference between the true and predicted time sequences is known as the \textit{innovation} sequence.  A successful prediction of leads to innovations that represent uncorrelated white noise.  (This is true for regression)%

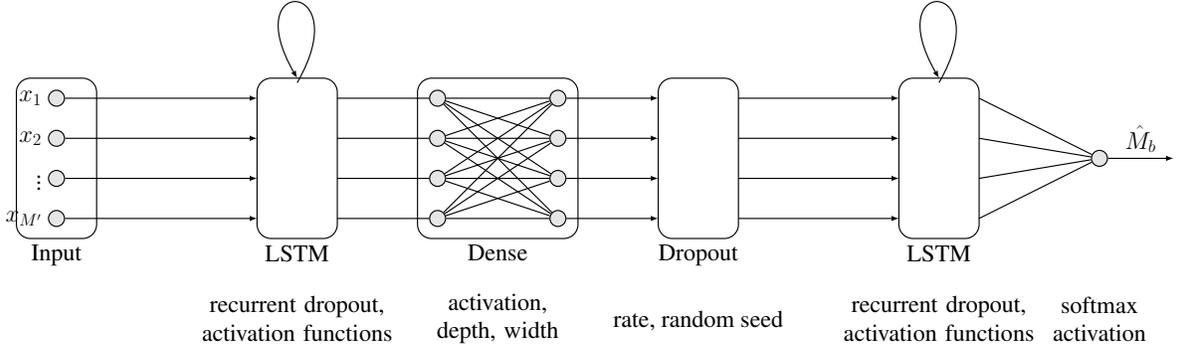
\begin{figure*}[!t]
%\begin{adjustwidth}{.25in}{0cm}
\centering
\resizebox{0.95\textwidth}{!}{\input{figures/lstm.tikz}}%
%\end{adjustwidth}
\caption{Architecture of the deep learner comprising fully connected deep neural networks, dropout layer, and LSTMs.}
\label{fig:lstm}
\end{figure*}%

\vspace*{-1em}
\subsection{Deep Learning}
Deep learning is a specific type of machine learning that uses several layers of connected computational elements called ``perceptrons,'' which are  threshold functions with weights and are the basic unit of a neural network.  These units can either be 1) connected where information flows forward between layers (i.e., from the input layer to the intermediate layer to to the output layer) or 2) recurrent where information can also flow through feedback connections.  As any machine learning algorithm, deep learning also aims at minimizing a cost function.  This is done through computing the gradient with respect to the weights of this function. %

To model temporal relationships between the various learning features, recurrent neural networks (RNNs) connect each time step with the previous ones through feedback connections.  However, RNNs suffer the vanishing (and exploding) gradient problems, whereby the gradient of the error function becomes vanishingly small (or explosively large) in the long-term, preventing the backpropagation algorithm from updating the weights.  This in turns prevents the RNNs from learning.  Another option is the use of LSTM that we propose.  Here, the computed weights can be ``forgotten'' thus keeping gradients unchanged.  A simplified solution is known as the gated recurrent unit (GRU).  GRUs simplify the forget gate operation through combining it with the input gate.  This effectively allows for faster training of smaller datasets at a compromise of learning longer time sequences.

Often mentioned alongside RNNs are convolutional neural networks (CNNs).  CNNs are useful in analyzing datasets that have grid patterns (e.g., classifying images or learning hierarchies in data) \cite{Goodfellow-et-al-2016}.  In comparison to RNNs (and by inheritence LSTMs), CNNs do not readily have temporal constructs.  Instead, CNNs have a property of spatial shift invariance, which essentially means that small location translations do not impact the performance of CNNs.  This is due to the pooling layer which effectively summarizes (i.e., through averaging or selection of extrema) the learning features.  In the case of a set of pre-defined trajectories, CNNs may not offer any additional performance improvement, since these trajectories are splines in two-dimensional space and are---by definition---summaries.%  Further, a trajectory is constructed due to motion over a period of time, and CNNs do not readily deal with time shifts, which require RNNs.

\textbf{Cost functions:} Are generally nonconvex functions of the true and predicted label vectors.  Despite using functions that are seemingly convex (e.g., cross-entropy), the use of hidden layers with non-linear activation functions can lead to several ``equal'' global minima making the function nonconvex.  As a result, an optimizer is needed to minimize the cost function. % https://stats.stackexchange.com/questions/144378/is-cross-entropy-cost-function-for-neural-network-convex

\textbf{Backpropagation:} In order for the information to flow between layers, the backpropagation algorithm computes the gradient of the cost function with respect to the weights of a perceptron one layer at a time.  Because of the nonconvexity of the loss function, a gradient-based optimizer is utilized to drive the prediction error to the lowest value possible.

\textbf{Tensors:} Tensors are generalizations of two-dimensional matrixes to a larger dimension space.  While the input data $\mathbf{\tilde D}$ is a two-dimensional time-indexed matrix, it has to be further reshaped to a tensor so that the third dimension is the time index\footnote{The order of this dimension does not have to be the third.  For example, with Keras \cite{chollet2015keras}, the time index is the first dimension, while the other two are simply a design matrix for each time index.}. % https://keras.io/api/layers/recurrent_layers/lstm/

\textbf{Activation functions:}  An activation function is a non-linear function\footnote{An exception is the linear activation function, which is used when the output data type is continuous.} that defines the output of a perceptron from a given set of inputs and weights.  Relevant ones in our paper are: 1) for LSTM, we use hyperbolic tangent as an activation for the memory cell state and sigmoid for recurrent input, forget, and output gates 2) for FCDNN, we use the rectified linear unit dense layers. The output layer of the model is a dense layer with a softmax activation function that calculates a probability for every possible class, which represents the unique beam identifier as motivated in \eqref{eq:beam_selection}.

\textbf{Initialization:}  For both LSTM and FCDNN we initialize the bias to zeros and the weights according to the Glorot algorithm, such that the variance of the activations are the same across every layer.  This helps prevent or minimize gradient saturation \cite{pmlr-v9-glorot10a}.  It an extensively used practice to optimize nonconvex functions though convergence may not be guaranteed \cite{Goodfellow-et-al-2016}.  

\textbf{Optimizer:}  To find the cost optimal weights, we use the adaptive moments (Adam) optimizer \cite{KingmaB14}. Adam has benefits of being computationally efficient, uses the second moment to accelerate the descent, with little memory being required.  Besides the learning rate which dictates the descent rate, other important parameters are 1) the batch size, which is the number of training samples in one optimizer step and 2) the number of epochs, which is the number of times the complete dataset is visited as part of the training phase.

\textbf{Hyperparameter tuning:}  Hyperparameters are deep neural network settings that are used to control its behavior.  To find the loss optimizing hyperparameter settings, we use grid search over a set of parameters related to the depth and width of the FCDNN.  This is done over data samples that the deep neural network does not observe as part of its training, referred to as the ``validation'' data. 

\textbf{Transfer learning:} Transfer learning \cite{5288526} is the process where a deep learner trained on one dataset is used 1) as a starting point for a model that performs a similar task or 2) to make inferences from a different and relevant dataset.  Thus, transfer learning is more time-efficient than training from scratch.  This becomes particularly useful in delay-sensitive RRM applications in wireless networks such as mobility. % https://www.mathworks.com/discovery/transfer-learning.html

\vspace*{-0.5em}
\subsection{Layers}
Here, we focus on the ones that are relevant to our paper: 1) fully connected deep neural network (FCDNN), 2) long short-term memory (LSTM), and 3) dropout layers.

\textbf{Fully connected deep neural networks} are made of several layers of artificial neural networks to approximate a function \cite{Goodfellow-et-al-2016}.  The first layer is the ``input layer'' while the last layer is referred to as the ``output layer.''  The hidden layers are all the layers between the input and the output layers.  In a FCDNN, the number of nodes in each hidden layer is its width and the number of the hidden layers is its depth.  FCDNNs do not factor in time relationships within the input data.

\textbf{Long short-term memory} are special types of RNNs.  RNNs are a family of neural networks that can process sequential data and thus are suitable for time-indexed datasets.  LSTMs comprise multiple elements: 1) memory cell state, which is generally common in RNNs 2) forget gate and 3) input gate \cite{lstm}.  The memory cell states simply record information.  The input gate is used to update the memory cell state. The forget gate can learn to reset the state of the memory cell when the stored information is no longer needed (i.e., invalidation of the memory cell).  If we denote vectors of the candidate memory cell (i.e., new information) as $\mathbf{\tilde c}_t$, the forget gate as $\mathbf{f}_t$, the input gate as $\mathbf{i}_t$, then we can represent the impact of the forget and input gate on the memory cell output as follows: $\mathbf{c}_t \coloneqq \mathbf{c}_{t-1} \odot \mathbf{f}_t + \mathbf{\tilde c}_t \odot \mathbf{i}_t$. 

\textbf{Dropout layer} randomly sets a fraction of the input units to zero at each training step, known as the \textit{rate}.  This effectively creates an ensemble of learners, which helps prevent overfitting (i.e., memorization of training data and failure of generalization).  The dropout layers exist in both the LSTM and FCDNN layers.

\textbf{Output layer} selects the class of the highest likelihood.  Since our deep learner is effectively a multi-class classifier, the softmax  activation function in the output layer converts the values from the to a vector of real values the sum of which is equal to 1. % Thus the softmax is a function $\sigma: \mathbb{R}^N \mapsto [0,1]^N$.

\textbf{Architecture} of the deep learner {with the relevant hyperparameters per building block is shown in} Fig.~\ref{fig:lstm}.  A high-level reasoning for this architecture is explained as follows: we start with an LSTM to learn long occurring patterns, then we use a fully connected deep neural network to extract additional learning features from these occurring patterns.  We drop some inputs at random to minimize overfitting and then we use another LSTM to learn any long term patterns from these additional features.  These learnings are fed to the output layer, which uses the softmax activation function.  The output of this function is a predicted class with the highest likelihood as motivated earlier.

\vspace*{-0.5em}
\subsection{Training and Validation Split}
Given that $\mathbf{D}$ is a time-indexed dataset, a random split between a training and a validation set often performed in supervised learning is not feasible.  This is because the implication of time and its impact on the values of each learning feature becomes overlooked.  Instead, a split that has two types of constraints that have to be fulfilled: 1) no randomness and 2) split must abide by the time boundaries of a radio frame.  To achieve these two constraints, the pivoting time index $t^\star$ at which the data is split is computed for a training data size $r_\text{training}$ as follows:

\begin{equation}
    t^\star \coloneqq \lfloor \lfloor r_\text{training} m^\prime \rfloor / N^\text{frame}_\text{slot} \rfloor  N^\text{frame}_\text{slot},
    \label{eq:data_size}
\end{equation}
where the discrete interval $[0,t^\star - 1]$ is the time indexes belonging to the training data while the interval $[t^\star, n - 1]$ is for the validation data.  Here $N^\text{frame}_\text{slot}$ is the number of time slots in a radio frame.

\subsection{Run-time Complexity}
\hl{Let us denote the LSTM input gate dimension as $I$ and the memory cell state dimension as $C$.  Further, let us denote the depth of the FCDNN as $d$ and the width of it as $w$.   Then, the run-time complexity of the architecture outlined in} Fig.~\ref{fig:lstm} \hl{as affected by the change in these values is in $\mathcal{O}(IC + C^2) + \mathcal{O}(wd)$} \cite{rnns, scikit-learn}, \hl{where the first term is the run-time complexity of the LSTM and the second term is the run-time complexity of the FCDNN, both as a function of their respective dimensions.}

% nX x width from line 225 in code.

% \textbf{Imbalance:} The number of classes (i.e., beam identifiers) are likely not equally likely in the dataset $\mathbf{D}$.  When classes are not represented equally, learning can be biased towards the majority class. Therefore, in order to enhance the learning, we oversample the minority classes using bootstrapping (i.e., random sampling with replacement). 
%%%%%%%%%%%%%%%%%%%%%%%%%%%%%%%%%%%%%%%%%%%%%%
\section{Beam Handoff Algorithms}\label{sec:beamhoffalg} %
%%%%%%%%%%%%%%%%%%%%%%%%%%%%%%%%%%%%%%%%%%%%%
%Here we discuss two sets of algorithms that apply to beam switching.
\subsection{Legacy}\label{subsection:legacy}
This follows the industry standards, where the network can configure UEs to measure and report the reference symbol for both serving and neighboring BS beams.  For RRC-based mobility, the network can configures events for mobility decisions such that if the SS-RSRP of the neighboring BS is above the serving BS by a certain offset and time-to-trigger duration (i.e., A3 event) \cite{3gpp38331}, the UE sends a measurement report to the network via the RRC protocol. %

% Alperen: these wre already motivated in 1A.  No need to rewrite here.
%\alperen{Based on the received measurements from the UE, the network sends an RRC message to trigger serving BS change.}
LLM on the other hand relies on beam-centric measurements where the network can configure UEs to periodically report serving and non-serving BS beams periodically \cite{R12102209}. %

%\alperen{The network can send MAC control element to trigger serving beam change based on the measurements.}

The minimum report interval for RRC-based mobility is $120$ ms \cite{3gpp38331} whereas the periodicity of the LLM beam reporting can be as low as $10$ ms \cite{38321}. %
%UE measures its received signal power on a beam (CSI-RS or SSB-RS) from a neighboring BS; if it is above the serving BS beam received signal power by a certain offset, the UE signals a handoff event to the serving BS.  The serving BS in turns facilitates the handoff procedure to the target BS.  In this case, at the time of the handoff completion, the UE will be served by the best beam as measured in the previous radio frame.
%{\color{blue} Event-based measurement reporting using event A3.  Minimum periodicity is 120ms (12 radio frames) XXXXXX
%CSI measurement reporting has a much smaller periodicity XXXXXX
%TODO: @alperen to add some details on CSI measurement reporting
%}

\hl{Because of the use of events and periodic reports, this algorithm does not benefit from any supervised learning and is essentially a reactive  (as opposed to a proactive) approach.}%

\subsection{Proposed}
%We use Layer 1 mobility during the training and then revert back to RRC reporting to conserve UE battery and reduce heat which can cause mmWave noise XXXXXXXX
In our proposed algorithm, we use both LLM and RRC-based mobility procedures.  The LLM procedure is adopted during the training for data collection where UEs report the SS-RSRP value per beam for several beams. There are other quantities construct the dataset and are either reported by the UE or computed at the BS as shown in Table~\ref{tab:quantities}.

\textbf{Objective:} The objectives tuple $\mathcal{O}$ are the radio measurements that the handoff aims at optimizing.  This could be the beam reference symbol received power or any other target as reported per UE for all time indexes.

\textbf{Trajectory:} We define a trajectory of a given UE as a time-indexed tuple for the $i$-th UE $\mathcal{T}_i(t) \coloneqq (t, i, \phi_i(t),b_i(t))$ collected over the period of movement of this UE in the association area of the network.  Here, the elements of this tuple are the time index, C-RNTI, direction ($\phi$) of the $i$-th UE, and beam identifier.  Optionally a time-indexed tuple can contain the longitude and latitude coordinates $(t, i, \phi_i(t), x_i(t), y_i(t), b_i(t))$.  The LSTM layers in the deep learner have an objective of learning these trajectories and the performance optimal beam as it changes over time over that trajectory.  The union of the trajectory and objective tuples at a given time index $t$ and for the $i$-th UE is essentially a row in the dataset $\mathbf{D}$.  We can formally define $\mathbf{D}$ \textit{before} the feature engineering steps as $\mathbf{D} \coloneqq [\mathbf{X}\,\vert\,\mathbf{y}] = [(\mathcal{T}_i(t) \cup \mathcal{O}_i(t))_{i=1}^u]_t$.

\textbf{Algorithm outline:} Algorithm~\ref{alg:algorithm} is outlined as follows:

\begin{enumerate}
    \item Collect UE LLM measurements and compute features as shown in Table~\ref{tab:quantities} to construct $\mathbf{D}$.  This contains the optimization objective measurement and helps construct trajectories of the UEs in motion.
    \item Train a deep learner and use its predictions of objective-optimal beams for all UEs following similar trajectories as learned by the deep learner:  If the predicted next optimal beam of the UE is different than the current serving beam, trigger a handover.
    \item If a UE hands off to a wrong beam or BS (i.e., as a result of a poor deep learner prediction), which could lead to a handover failure as UE may experience a weak signal, %\alperen{this can lead to handover failure as UE will not be able to connect to the target beam or BS due to weak signal strength of the target beam or BS}. 
    invoke fallback:
    {UE to initiate recovery procedures (e.g., RRC reestablishment or beam failure recovery).} This effectively pauses the LLM measurements.
    \item Perform transfer learning using new data (or \textit{experience}) obtained from either good or bad predictions.
    \item Resume LLM measurements.
    \item This outline repeats as long as the algorithm is enabled.
\end{enumerate}

Due to the predictions made by this algorithm, the UEs do not need to periodically report their radio measurements, which helps 1) conserve UE battery 2) reduce heat that is a major contributor to shot noise at mmWave frequencies and 3) reduce control signaling overhead.  The algorithm is also robust against failure due to the fallback mechanism: If the prediction performance of the network is poor, the UE can either indicate beam failure or a RLF.  Then, the UE and initiates an RRC re-establishment or beam failure recovery procedure to connect to the network. The proactive handoff for this UE is stopped until the re-establishment of the connection between UE and the network.

%, if the UE ends up handing off to the wrong beam, a RLF event happens and an RRC procedure will be triggered for the UE to handoff to a feasible candidate.  This RRC procedure is also a trigger for the transfer learning to take place and for the network to retrain using more LLM measurements.

\SetKwFor{Loop}{Loop}{}{EndLoop}
\begin{algorithm}[!t]
    % \small
    %\renewcommand\thealgorithm{} % use to suppress the algorithm number
    \caption{Proactive Optimal Beam Handoff}
    \label{alg:algorithm}
    \DontPrintSemicolon
    \KwIn{Optimization objective from $\mathcal{O}$, training-validation ratio from \eqref{eq:data_size}.}
    \KwOut{Objective-optimal beam assignment to the $i$-th UE for the next radio frame for all $i\in\{1,2,\ldots,u\}$.}
    \Loop{} {
        \If {moving UEs are in LLM} {
            Collect reports and computations to construct $\mathbf{D}$ from the features shown in Table~\ref{tab:quantities}. \;
            Train a deep learner using first \eqref{eq:data_size} samples and assess prediction performance: \;
            \eIf {validation performance is acceptable} {
                Use deep learner predictions of optimal beams based on the trajectories as learned by the deep learner. \;
                Proactively handoff UEs to their predicted optimal beams if the predicted next best beam is different than the current serving beam. \;
                If UEs report un-degraded measurements (at least), set prediction to good. \;
            }
            {
                Set prediction to bad.\;
            }
        }
        \If {\text{\rm prediction is bad}} {
            Fallback: UE initiates recovery procedure (e.g., RRC re-establishment or beam failure recovery). \;
            %UE initiates a RRC procedure for either a radio link failure (RLF) or RRC mobility. \;
            Instruct UEs to pause LLM measurements. \;
        }
        Retrain deep learner through transfer learning (if applicable) using recent experience from fallback. \;
        Instruct UEs to resume LLM measurements. \;
    }
\end{algorithm}%

\begin{table*}[!t]
\centering
\caption{Dataset Features}
\vspace*{-1em}
\label{tab:quantities}
\begin{threeparttable}
\begin{tabular}{ p{20em}ll } 
%\hhline{===}
\hline
Parameter & Data type & Owner \\
\hline
Radio frame number & Integer & Computed by the BS \\
C-RNTI & Integer & Assigned by BS to the UE \\
Current beam index  & Integer &  Reported by the UE\tnote{\textasteriskcentered} \\
Previous beam index  & Integer &  Reported by the UE\tnote{\textasteriskcentered}\\
Beam SS-RSRP & Float & Reported by the UE \\
%Beam SINR & Float & Reported by the UE \\
UE direction & Integer & Computed by the BS \\
UE speed & Float & Computed by the BS \\
UE coordinates & Float & Reported by the UE \\
Radio link failure indicator & Boolean & Reported by the UE   \\
%\hhline{===}
\hline
\end{tabular}
\begin{tablenotes}\footnotesize
\item[\textasteriskcentered]{These are assigned by the serving BS.}
\end{tablenotes}
\end{threeparttable}
\end{table*}%

\textbf{Training approaches:} We propose two methods for model training: 1) Distributed and 2) Centralized.

% NOTE: Faris says: We will not have no fading based on our discussions on 7/7
\subsubsection{Distributed} In this case a deep learner is created per BS-UE pair.  Thus, the UE identifier is necessary and transfer learning \textit{cannot} be applied since the C-RNTI can be reassigned to a different UE \cite{3gpp38331}.  This allows us to come up with two approaches of the distributed model:

\textit{Proposed with UE coordinates:} In this scheme we allow the UEs to report its coordinates to the serving BS, which uses the coordinates as learning features to predict the optimal beam for these UEs.

\textit{Proposed without UE coordinates:} UEs do not report their coordinates $(x_i,y_i)$ to their serving BS.

In the distributed case, the number of deep learners required is $u\vert\mathcal{B}\vert$.  This is because one deep learner is assigned per \mbox{BS-UE} pair.

\subsubsection{Centralized} In this case a deep learner is created per BS and the BS does not require any knowledge about the UE identifier or positions.  Instead, it ``crowdsources'' knowledge regardless of which UE reports it.  This clearly reduces the number of deep learners required to $\vert\mathcal{B}\vert$. % and the number of columns in $\mathbf{D}$ to XXX

%Another approach to centralization can be using a single deep learner for all the UEs and BSs in one geographical area.  The details of this approach and its algorithmic implications are intended for future work.

%%%%%%%%%%%%%%%%%%%%%%%%%%%%%%%%%%%%%%%%%%%%%
\vspace*{-0.5em}
\section{Simulation}\label{sec:simulation}%
%%%%%%%%%%%%%%%%%%%%%%%%%%%%%%%%%%%%%%%%%%%%%
\subsection{Setup}
\begin{figure*}[!t]
%\begin{adjustwidth}{.25in}{0cm}
\centering
\resizebox{0.75\textwidth}{!}{\includegraphics{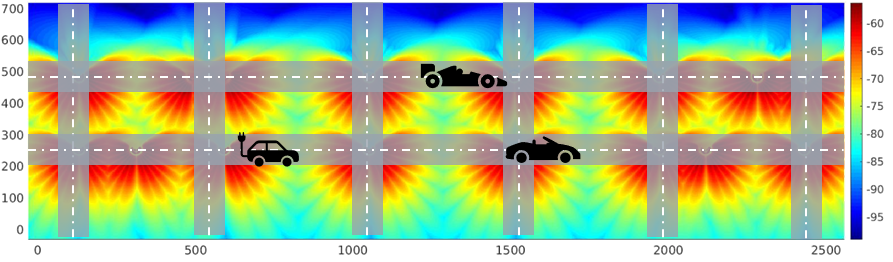}}%
%\end{adjustwidth}
\caption{Manhattan grid.  Base stations are uniformly distributed at each road intersection.  The received signal power per beam [dBm] is displayed for all beam candidates.}
\label{fig:manhattan}
\end{figure*}

%\subsubsection{Data Generation}
The dataset used for the simulation is generated from a system level simulator based on the Manhattan grid as visualized in Fig.~\ref{fig:manhattan}.  This grid consists of two streets the width (and height) of each is $200$ m (and $100$ m). These distances are influenced by the empirical measurements from \cite{7481506}.  The radio propagation in this scenario provides street canyon like effects and sharp transitions between line of sight (LOS) and non-LOS. We use micro BSs each of which has an antenna height of $5$ m. % There are $32$ antenna elements, so we compute $\Theta$ from \eqref{eq:coherence} as $\Theta \approx 102/32 \approx 3.2$ radians. % 
%The inter-site distance varies from $100$ m to $460$ m.
UEs are independently and uniformly distributed in the grid and move in one of four directions at random (uniform) and at an average speed of $100$ km/h. We use full buffer traffic model (i.e., infinite data availability) to keep the system at maximum load.  To add an element of complexity, the speed of the UEs changes every $100$ m with probability of $0.2$. We set $r_\text{training}$ to $0.6$ and run the simulation for a total of $50$ seconds or $5{,}000$ radio frames.  Further details of the simulation setup can be found in Table~\ref{tab:simulation}.%

%We consider a Manhattan grid with XXXXXX as in Fig.~\ref{fig:manhattan}.  The simulation parameters are show in in Table~\ref{tab:simulation}

\begin{figure}[!t]
%\begin{adjustwidth}{.25in}{0cm}
\centering
\resizebox{0.5\textwidth}{!}{\input{figures/BS0att_1_using_7history_frames_for_MT_11.tikz}}%
%\end{adjustwidth}
\caption{Accuracy and loss for both the training and validation data for a given BS-UE pair and $\ell = 7$ lookback.}
\label{fig:loss_vs_epoch}
\end{figure}
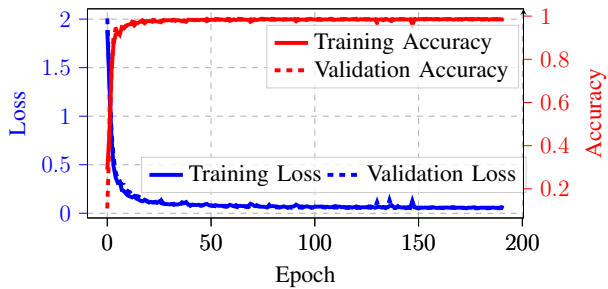

% The UEs measure the RSRP values of the SSB-RS beams of the BSs and shares with the network for data collection with the periodicity of 10 ms. The training of the model for handover prediction is initiated once enough measurements from the various UEs is collected. In total, there are 400 UEs e.g. ~25 UEs per site in the simulation and the simulation is run for 50 secs.

%\subsubsection{Training}
%The models are trained  192 epochs with batch size of 64. 
The architecture of the deep learner is depicted in Fig.~\ref{fig:lstm}.  The deep learner has depth of $8$ and width of $16$.  It is trained with $192$ epochs and a batch size of $64$, each with a rectified linear unit activation function.    The data training-validation split is 60\%-40\%.  A dropout rate of 0.2 is used for the dropout layer.  We also set the random generator seed to $0$ for reproducibility.

We run three flavors of the proposed algorithm: 1) distributed with UE coordinates, 2) distributed without UE coordinates, and 3) centralized.  The performance of these is benchmarked against a perfect predictor.  The objective of interest is the SS-RSRP.  That is, we proactively instruct UEs to handoff to received power-optimal beams as predicted by the next radio frame.  For all three flavors (or approaches), we choose a lookback $\ell$ from $\mathcal{L} \coloneqq \{0,1,\ldots,10\}$.  Centralized approach ignores two features from: 1) C-RNTIs (which are used in the distributed approach) and 2) the UE coordinates.%

\subsection{Results}
% The Figure \ref{fig:msg1_dist_cent_lb_acc} shows the distribution of the lookback window that provides the highest prediction accuracy for each BS for distributed and centralized approach. The average accuracy of the BS performance is calculated based on the mean of accuracy for each UE in the distributed approach. It can be seen that higher lookback window for training provides better accuracy for distributed approach e.g. the ~30\% of the best predictions is performed with lookback window 10. The behaviour of the lookback frame impact is not stable for centralized approach but 7 or more lookback frames provides better performance.

We start this subsection with a few important definitions related to the performance of the proactive beam handoff algorithm.  These are the accuracy and the relative accuracy.

\textbf{Average accuracy:} The average accuracy for a multi-class classification can be challenged as a suitable measure since the beam idenitifiers may not be balanced in their count.  However, what the UE is interested in a proactive handoff is whether the predicted and true beam identifier are identical.  Thus, the accuracy ($\bar A$) is defined as an average of identical values of the classes:
\begin{equation}\label{eq:accuracy}
    \bar A(\mathbf{y}, \mathbf{\hat y}) \coloneqq \frac{1}{V} \sum_{i=0}^{V-1} \mathbbm{1} \left [ [\mathbf{y}]_i = [\mathbf{\hat y}]_i \right ],
\end{equation}%
where $V$ is the dimension of the beam identifier vector, collected over the validation time period \eqref{eq:data_size} for all UEs.  The average accuracy {is a significant measure for radio access performance}
%has a radio access performance significance: 
{since a mispredicted beam results in wrong handoff decision and thus can degrade quality of experience and lead to mobility failure or increase the duration of service interruption \cite{9417452}.}  Also, the RLF rate due to beam handoff failure is simply $1 - \bar A$.

\textbf{Zero-one score:} To assess the added value of the UE position to the prediction performance, we define a vector of loss-win scores known as the zero-one score ($\boldsymbol\alpha \in \{0,1\}^V$) as:
\begin{equation}
    \boldsymbol\alpha(\mathbf{\hat s}, \mathbf{\hat s_\text{base}}; \varepsilon) \coloneqq \mathbbm{1} \left [ (\mathbf{\hat s} - \mathbf{\hat s_\text{base}} ) > \varepsilon\mathbf{1} \right ],
\end{equation}%
where the $V$-dimensional column vectors are for a certain performance measure (e.g., average accuracy), $\varepsilon$ is the cutoff threshold, and $\mathbbm{1}[\cdot]$ is overridden as the indicator \textit{vector}. % set \varepsilon to 0.03

% %\input{figures/lookback_s1s2.tikz}
% \begin{figure}[!t]
% %\begin{adjustwidth}{.25in}{0cm}
% \centering
% \resizebox{0.5\textwidth}{!}{\input{figures/msg1_dist_cent_best_predictionvslookback.tikz}}%
% %\end{adjustwidth}
% \caption{distributed vs centralized approach best accuracy vs lookback}
% \label{fig:msg1_dist_cent_lb_acc}
% \end{figure}

For the results, we begin with the loss \eqref{eq:loss} as shown in Fig.~\ref{fig:loss_vs_epoch}.  As the number of epochs increase we observe a decrease in the loss and increase in the accuracy which shows that the deep learner is indeed ``learning'' from $\mathbf{D}$.  This demonstrates that proactive beam handoff using deep learning is possible.  

\begin{figure}[!t]
%\begin{adjustwidth}{.25in}{0cm}
\centering
\resizebox{0.5\textwidth}{!}{\input{figures/msg1_distvscent_average_accuracy.tikz}}%
%\end{adjustwidth}
\caption{The average accuracy of the proactive beam handoff approach vs.\ lookback.}
\label{fig:msg1_dist_cent_lb_accv2}
\end{figure}
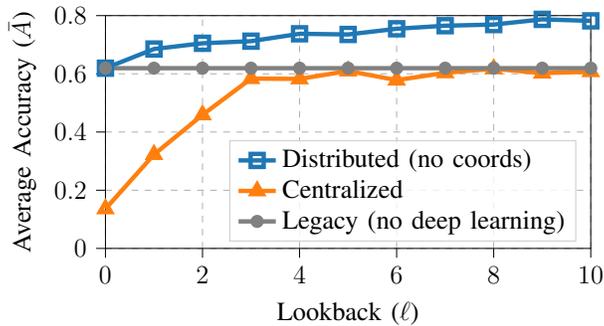

In Fig.~\ref{fig:msg1_dist_cent_lb_accv2}, we show the average accuracy of all legacy, distributed, and centralized approaches over several lookback values.  The performance of the distributed approach to proactive handoff is superior to the centralized one.  With $\ell = 0$, the distributed case with no coordinates is equivalent to the legacy approach outlined in Section~\ref{subsection:legacy}. We discuss why the performance of the distributed approach is superior to the centralized one in the next subsection. 

\begin{figure}[!t]
%\begin{adjustwidth}{.25in}{0cm}
\centering
\resizebox{0.5\textwidth}{!}{\input{figures/msg2_acc_.tikz}}%. % msg2_accuracy_dist_dist_pos_lb0_10_rmBS0.tikz}}%
%\end{adjustwidth}
\caption{Vector plot of $\boldsymbol\alpha$ of the difference between impact of UE positions compared to without UE positions for the distributed model.}
\label{fig:msg2_distPos_dist_acc}
\end{figure}
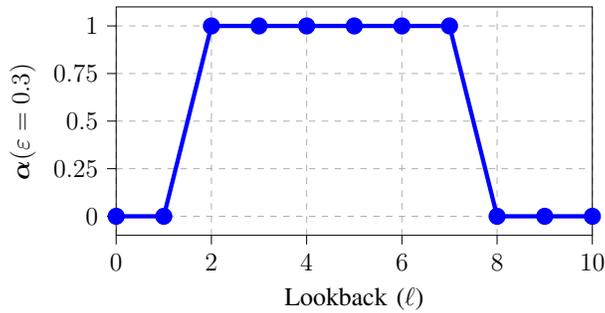

Fig.~\ref{fig:msg2_distPos_dist_acc} shows the zero-one score with parameter $\varepsilon = 0.03$.  Here, the baseline is the performance of the distributed approach without UE coordinates.  We observe that the distributed approach using the UE coordinates does not always outperform baseline. Particularly, we observe that the deep learner does not perform well with short lookback sequences, which is expected as LSTM generally penalizes shorter sequences.  However, when the number of lookbacks is between $2$ and $7$, then the distributed approach with UE coordinates outperforms the baseline.  We also notice that for larger $\ell$, knowledge of UE coordinates does not bring any additional insight to the deep learner.  Further details are discussed in the next subsection.

%It can be seen that if the network has only certain number of previous measurements from the UE e.g. [2,7], having position information increases the training accuracy.

\begin{figure}[!t]
%\begin{adjustwidth}{.25in}{0cm}
\centering
%\resizebox{0.5\textwidth}{!}{\input{figures/msg3_dist_pos_lb0_10_acc_BS.tikz}}%
\resizebox{0.5\textwidth}{!}{\input{figures/msg3_dist_faris_excel.tikz}}%
%\end{adjustwidth}
\caption{The maximum, average, and minimum prediction accuracy of the distributed proactive beam handoff algorithm with various lookbacks.}% \hl{and with the algorithm turned off}.}
\label{fig:msg3_BSvs_acc}
\end{figure}
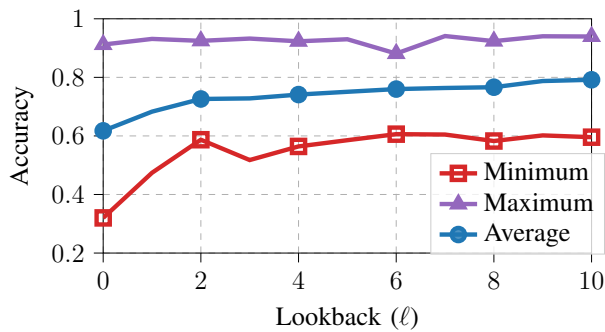

To understand the extent of the performance of the distributed approach using the UE coordinates, we plot the aggregated accuracy values for all BS-UE pairs grouped by the lookback values in Fig.~\ref{fig:msg3_BSvs_acc}. The takeaway is that the performance of the model varies across different users in the network and that not all users would similarly benefit from a proactive handoff. % XXXXX varies depending on the BS deployment. One can analyze the performance differences among BSs with system performance of the BSs in terms of average BS SINR, throughput, interference, CQI, etc. etc. in future works.  Ozge's comment about after a certain number of lookbacks, there is no advantage for the knowledge of the UE coordinates. XXXXXX

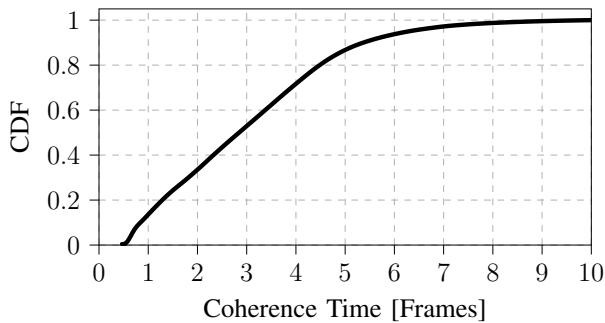
\begin{figure}[!t]
%\begin{adjustwidth}{.25in}{0cm}
\centering
\resizebox{0.5\textwidth}{!}{\input{figures/coherence.tikz}}%
%\end{adjustwidth}
\caption{Channel coherence cumulative distribution function (CDF).}
\label{fig:coherence_time}
\end{figure}%

\begin{table*}[!t]
\centering

\caption{Simulation settings}
\vspace*{-1em}
\label{tab:simulation}
\begin{tabular}{ p{20em}l } 
\hline
%Center frequency & $28$ GHz\\
%Subcarrier spacing & 15 KHz \\
%Carrier bandwidth & $20$ MHz \\
Technology and duplex mode & 5G TDD \\
Channel model & 3GPP TR 38.901 Urban Micro \cite{3gpp38901}\\
Beams per base station (BS) & $8$ \\
BS azimuth beam angles &  $\{44^\circ, 56^\circ, 69^\circ, 82^\circ, 96^\circ, 109^\circ, 122^\circ, 134^\circ\}$ \\
Traffic model & Full Buffer \\
Timeslots per radio frame & $10$ \\
%Control and RS overhead & 21\% \\
BS transmit power & $44$ dBm \\
Number of BSs ($\vert\mathcal{B}\vert$) & $32$ \\
Number of UEs ($u$) & $400$ \\
%\alperen{UE antenna configuration} & \alperen{ 2RX, 2TX isotropic }\\
%UE receiver & LMMSE-IRC \\
%UE receiver noise figure & 9 dB \\
\hline
\end{tabular}
\end{table*}%

% Faris says: we are already doing this.  A multi-class classifier is effectively multiple binary classifiers.
% An observation from Fig.~\ref{fig:msg3_BSvs_acc} motivates developing higher granularity approaches where a model can be specialized for each beam pair between UE and serving beams in order to improve the performance.  This approach has a high computational overhead, which may be prohibitive to its implementation, but is an interesting extension to the distributed approach.
% % Goes to discussion

\subsection{Discussion}
We start with discussing the performance disparity in both the centralized and distributed approaches with no UE coordinates in Fig.~\ref{fig:msg1_dist_cent_lb_accv2}.   This  disparity is attributed to a wide range of trajectory-dependent measurement values that several UEs simultaneously take.  In other words, a given point on the trajectory can have several different measurements depending on which serving beam they were on. % 
%\alperen{XXXXX I think, there can be other factors that affects UE-beam pair measurements although the UE is in the same coordinates such as interference caused by other UEs around, randomness in blockage due to dynamic environment e.g. mobility of other agents, etc. }
This applies to all points on the trajectory alike.  The absence of the C-RNTI drives the centralized deep learner (one learner per BS) to learn a weighted average of measurements which prevents accurate prediction of the objective-optimal beam identifier.  However, with the distributed approach, the C-RNTI allows these trajectory-dependent measurements to be uniquely attributed the UE reporting them, causing the learner to learn unique measurements along the trajectory---a task that an LSTM is designed to handle.  
\hl{A natural question after this discussion is: How do these two algorithms compare with the legacy algorithm where no proactive beam handoff exists? To answer this, we remark that the legacy algorithm is a per BS-UE algorithm.  This is because each UE either reports an RRC event or a periodic beam measurement independently from other UEs, and the BS decides whether or not a beam handoff should take place.   Thus, the comparison with the distributed method of our proposed algorithm is more suitable (which is also BS-UE pair). It is clear that the legacy algorithm does not depend on the number of lookbacks since it depends on the most recent measurement only.  We observe that the legacy approach underperforms the distributed proposed approach.  One reason why this happens is because the knowledge of the trajectory and its underlying radio conditions during the handoff enables the proactive approach to assign a more robust beam (e.g., compared to a beam with stronger SS-RSRP that is short-lived in the trajectory that the legacy algorithm assigns).}

Next, we study the impact of the knowledge of UE coordinates in the proactive beam handoff.  We do this through studying the impact of lookback on the zero-one score as shown in Fig.~\ref{fig:msg2_distPos_dist_acc}.  Smaller sequences could drive LSTMs towards over-triggering their forget gates, which deletes the information in the self-recurrent memory BSs. % https://d2l.ai/chapter_recurrent-modern/seq2seq.html
However, as the number of sequences grown larger, we observe two important behaviors which can be mapped to two different regimes:
\begin{enumerate}
    \item Short lookback regime: any number of lookbacks below $7$ (but greater than $1$) radio frames.
    \item Long lookback regime: any number of lookbacks greater than $7$ radio frames.
\end{enumerate}

An important question can be asked here: What is the significance of such a breakdown?  To answer this question, we observe that the win-loss score for the distributed approach using the UE coordinates versus the approach not using them is equal to $0$ after $\ell = 7$.  This is because of the capability of LSTM to learn long sequences.  As a result, the knowledge of UE coordinates is no longer relevant since the deep learner can implicitly learn them with high accuracy through the beams and the time-based trajectory.  At $\ell = 7$, we have a total of $n = 2(7 + 1 + 1) = 18$ learning features for $\mathbf{D}$.  We cannot extrapolate $\ell$ indefinitely as $\ell$ cannot exceed the beam coherence time \eqref{eq:coherence}.  We show the coherence time for the UEs in Fig.~\ref{fig:coherence_time}, where only $1.2\%$ of the samples represent coherence times greater than $7$ radio frames.  This is expected for high-mobility UEs.

Finally, we notice that not all UEs benefit from the proactive beam handoff equally.  Fig.~\ref{fig:msg3_BSvs_acc} shows that while some UEs have an accuracy consistently above $0.9$ for any lookback value, there are some UEs that do not enjoy that.  This depends on the proximity of these UEs to their serving BS %
%\alperen{XXXX I think, it can be an interesting future direction to investigate in the context of explainable AI e.g. what are the factors that leads to different behaviour for different UEs.}\faris{XXXX This can be a separate paper but not necessarily disclosed.  We should limit the number of future directions otherwise it will weaken our contribution.}
and the uniqueness of the trajectories they follow.  Based on the Manhattan grid in Fig.~\ref{fig:manhattan}, we compute that the inter-site distance can be as short as $100$ m and as long as $460$ m.  If trajectories have small number of overlaps, then LSTMs can learn the trajectory and the optimal beams on its path.  %
% \alperen{XXXX We can generate a CDF/PDF of accuracy based on the BS-UE distance to understand this hypothesis better? } \faris{XXXX I don't understand what is a CDF/PDF of accuracy based on distance?  Are you referring to a joint PDF?} \alperen{what I had in mind is that, we can create bins between e.g. 0-400mt with granularity of 10m, where 0 the minimum distance between UE-serving BS and 400m is the max distance between UE and BS in the simulation. For each prediction, we know the distance between UE and the serving BS, and whether it was correct or wrong. So, we can categorize the predictions based on the distance between UE and BS, then calculate the accuracy of predictions for each interval, the generate PDF/CDF where X axis is the distance between UE-serving BS.} \faris{XXXX I see.  This is not technically a PDF or a CDF; however, it is not possible unless we run the entire simulation again!  Current data results do not have this info.}
However, if UEs are on trajectories that overlap with other trajectories, such as turns, then undesired handoff effects such as ping-pong could cause conflicting data to be added to the dataset.  This would cause the LSTMs to fail in learning the optimal beams.  If beams are equally likely to serve UEs, then the lower bound of accuracy is $0.125$, which is the reciprocal of the number of beams $8$, assuming equally likely occurrence of beams. %
%Show plots and performance based on intra-BS vs inter-BS.  Alperen will check if FREAC data shows both flavors PrevBS and CurrBS.

%%%%%%%%%%%%%%%%%%%%%%%%%%%%%%%%%%%%%%%%%%%%%
\section{Conclusion}\label{sec:conclusion}
%%%%%%%%%%%%%%%%%%%%%%%%%%%%%%%%%%%%%%%%%%%%%
In this paper, we demonstrated that proactive beam handoff is possible using deep learners with deep learners created per BS-UE pair residing at the edge.  We showed three different approaches.  The introduction of the UE coordinates help enhance the performance of the deep learner towards the proactive beam handoff; however as the longer sequences are presented to the deep learner, the need for the UE coordinates becomes diminished.  The use of intelligence in radio resource management algorithms help reduce the dependency on the UEs therefore preserving their battery life and reducing their temperature, which helps combat shot noise---a major issue in mmWave communications.

\bibliographystyle{IEEEtran}
\bibliography{main.bib}

\end{document}

%% file: figures/overall.tikz
\begin{tikzpicture}[style=thick, node distance=2cm, scale=1.2, >=latex] % change to 1 for two column
    \node [coordinate, name=input] {};
	\node [rectangle, draw, right of=input, 
		text width=10em, text centered, minimum height=4em] (BS) {Next-Generation Base Station};
    
    \node [rectangle, draw, below of=BS, node distance=2cm,
 		text width=10em, text centered, minimum height=2em] (RIP) {Per-Beam Radio Measurements};	
    
    \node [rectangle, draw, left of=RIP, node distance=5.5cm, xshift=-1cm, % change to 3.5 for two column
		text width=8em, text centered, minimum height=2em] (RIC) {Edge Node};

    \node [rectangle, fill=gray!40, draw, above of=RIC, node distance=3em, yshift=4em,
        text width=10em, text centered, minimum height=2em] (unsup) {Beam Prediction};

    \draw [draw,->] (RIP.180) -- node[yshift=-0.7em] {\small Storage} (RIC.0);
    \draw [draw,<-] (BS.270) -- node {} (RIP.90);
    \draw [draw,->,dashed, line width=0.5mm] (RIC.90) -- node[yshift=1.2em, xshift=1.75em] {\small Compute} (unsup.270);
    \node [rectangle, draw, fill=white, opacity=1, above of=RIC, node distance=3em,
		text width=10em, text centered, minimum height=2em] (AI) {Deep Learning};

\end{tikzpicture}%

%% file: figures/signaling_assign.tikz
\begin{tikzpicture}[node distance=6cm, scale=1, style=thick, auto] % change scale back to 1 for two column
\tikzstyle{every node}=[font=\small]

% Empty ladder diagram
\node[draw, rectangle] (UE)  at (0,0) {UE};
\node[draw, rectangle, right of=UE] (NR) {BS};
\foreach \x in {0,6}
    \draw (\x,-0.25) -- (\x,-3.5);

% Signaling flow here
\draw[->,>=latex] (0,-1)  -- node [text width=6cm, align=center, above] {MSG1: Random Access Request} ++(6,0);
%\node[circle,fill=black,inner sep=0pt,minimum size=5pt,label=right:{A}] (a) at (3,-1) {};
\draw[<-, >=latex, line width=0.35mm] (0,-3)  -- node [text width=6cm, align=center, above] {MSG2: Random Access Response \\ C-RNTI assignment} ++(6,0);
%\node[rectangle,fill=black,inner sep=0pt,minimum size=5pt,label=right:{D}] (d) at (3,-2) {};
% \draw[<-, >=latex] (0,-3)  -- node [text width=4cm, align=center, midway,above] {\scriptsize{RRC Mobility From E-UTRA Command}} ++(3,0);
% \draw[<->, >=latex] (0,-3.5)  -- node [text width=4cm, align=center, midway,above] {\scriptsize{Random Access}} ++(3,0);
%\node[circle,fill=black,inner sep=0pt,minimum size=5pt,label=right:{B}] (b) at (3,-3) {};
\end{tikzpicture}

%% file: figures/lstm.tikz
\begin{tikzpicture}[thick, scale=1, >=latex, font=\Large, cnode/.style={draw=black,fill=#1,minimum width=3mm,circle}]

\setstretch{1.0}
    %%%%%%%%%%%%%%%%%%%%%%%%%%%%%%%%%%%%%%%%%%%%%%%%%%%%%%%%
    % Blocks
    \draw[rounded corners=10pt] (0,-1) rectangle ++(2,-4) node[below] at (1,-5)  {Input};
    \draw[rounded corners=10pt] (6,-1) rectangle ++(2,-4) node[below] (LSTM1) at (7,-5)  {LSTM};
    \draw[rounded corners=10pt] (10,-1) rectangle ++(4,-4) node[below] (DNN) at (12,-5) {Dense};
    \draw[rounded corners=10pt] (16,-1) rectangle ++(2,-4) node[below] (Dropout) at (17,-5)  {Dropout};
    \draw[rounded corners=10pt] (22,-1) rectangle ++(2,-4) node[below] (LSTM2) at (23,-5)  {LSTM};    
    
    \node[cnode=gray!20] (output_3) at (27, -3) {};
    \node [coordinate, right of=output_3, xshift=2em, name=output_4] {};
   	\draw [draw,->] (output_3) -- node[above] {$\hat M_b$} (output_4);
   	%%%%%%%%%%%%%%%%%%%%%%%%%%%%%%%%%%%%%%%%%%%%%%%%%%%%%%%%
   	
   	% Generate recurrent loops to distinguish LSTM
   	\path[->] (LSTM1) +(0,4.25) edge [in=120,out=60,distance=3cm,loop,looseness=1] (7,-1); % node {$x$} (LSTM1)
   	\path[->] (LSTM2) +(0,4.25) edge [in=120,out=60,distance=3cm,loop,looseness=1] (23,-1);
   	%%%%%%%%%%%%%%%%%%%%%%%%%%%%%%%%%%%%%%%%%%%%%%%%%%%%%%%%
    % The input
    \foreach \x in {1,2,...,4}
    {
      \pgfmathparse{\x== 3 ? "\vdots" : "x_\x"}
             \pgfmathparse{\x== 4? "x_{M^\prime}" : "\pgfmathresult"}
        \node[cnode=gray!20,label=180:${\pgfmathresult}$] (x-\x) at (1,{-\x-int(\x/10)-0.5}) {};
        % Connections from input to lSTM
        \draw [draw,->] (x-\x) -- ++(5,0);
    }
    %%%%%%%%%%%%%%%%%%%%%%%%%%%%%%%%%%%%%%%%%%%%%%%%%%%%%%%%
    
    %%%%%%%%%%%%%%%%%%%%%%%%%%%%%%%%%%%%%%%%%%%%%%%%%%%%%%%%
    % The dense
    \foreach \x in {1,2,...,4}
    {
        \node[cnode=gray!20] (din-\x) at (10.5,{-\x-int(\x/10)-0.5}) {};
        \node[cnode=gray!20] (dout-\x) at (13.5,{-\x-int(\x/10)-0.5}) {};
    }
    
    % the connections between dense
    \foreach \x in {1,...,4}
    {   
        \foreach \y in {1,...,4}
        {  
	        \draw (din-\x) -- (dout-\y);
        }
        
        % Connections from dense to dropout
        \draw [draw,->] (dout-\x) -- ++(2.5,0);
    }

    %%%%%%%%%%%%%%%%%%%%%%%%%%%%%%%%%%%%%%%%%%%%%%%%%%%%%%%%
    
 	%%%%%%%%%%%%%%%%%%%%%%%%%%%%%%%%%%%%%%%%%%%%%%%%%%%%%%%%
 	% The coordinates from 2nd LSTM to output
 	\foreach \x in {1,2,...,4}
    {
        \node [coordinate] (c-\x) at (24,{-\x-int(\x/10)-0.5}) {};
    }
    % the connections from 2nd LSTM to output
    \foreach \x in {1,2,...,4}
    {   
        {   \draw (c-\x) -- (output_3);
        }
    }
    %%%%%%%%%%%%%%%%%%%%%%%%%%%%%%%%%%%%%%%%%%%%%%%%%%%%%%%%
    
 	%%%%%%%%%%%%%%%%%%%%%%%%%%%%%%%%%%%%%%%%%%%%%%%%%%%%%%%%
 	% The coordinates from Dense to dropout
 	\foreach \x in {1,2,...,4}
    {
        \node [coordinate] (l1-\x) at (7,{-\x-int(\x/10)-0.5}) {};
        \node [coordinate] (l2-\x) at (10,{-\x-int(\x/10)-0.5}) {};
    }
    % the connections from dense to dropout
    \foreach \x in {1,2,...,4}
    {   
        \draw (l1-\x) ++(1,0) -- (din-\x); % LSTM1 to Dense
        \draw[draw,->] (dout-\x) ++(4.5,0) -- ++(4,0); % from Dropout to LSTM
        
    }
    %%%%%%%%%%%%%%%%%%%%%%%%%%%%%%%%%%%%%%%%%%%%%%%%%%%%%%%%

    % Now write under each LSTM
    \node[below of=LSTM1, yshift=-1.5em, align=center, text width=12em] {\review{recurrent dropout, \\ activation functions}};
    \node[below of=DNN, yshift=-1.5em, align=center, text width=10em] {\review{activation, depth, width}};
    \node[below of=Dropout, yshift=-1.5em] {\review{rate, random seed}};
    \node[below of=LSTM2, yshift=-1.5em, align=center, text width=12em] {\review{recurrent dropout, \\ activation functions}};
    
    \node[below of=output_3, yshift=-7.125em, align=center, text width=8em] {\review{softmax activation}};

\end{tikzpicture}

%% file: figures/BS0att_1_using_7history_frames_for_MT_11.tikz
% This file was created with tikzplotlib v0.10.1.
\begin{tikzpicture}

\definecolor{darkgray176}{RGB}{176,176,176}
\definecolor{lightgray204}{RGB}{204,204,204}

\begin{axis}[
width=3.7in,
height=2.1in,
tick align=outside,
legend cell align={left},
legend columns=4,
legend style={
  fill opacity=0.8,
  draw opacity=1,
  text opacity=1,
  at={(0.565,0.12)},
  anchor=south,
  draw=lightgray204
},
tick pos=left,
x grid style={darkgray176,dashed},
xlabel={Epoch},
xmajorgrids,
xmin=-9.55, xmax=200.55,
xminorgrids,
xtick style={color=black},
y grid style={darkgray176,dashed},
ylabel={\textcolor{blue}{Loss}},
ymajorgrids,
ymin=-0.0485290423035622, ymax=2.10610268861055,
yminorgrids,
ytick style={color=blue},
yticklabel style={color=blue}
]
\addplot [line width=2pt, blue]
table {%
0 1.88747048377991
1 1.34685921669006
2 0.788471698760986
3 0.489103674888611
4 0.371434807777405
5 0.315249443054199
6 0.326119750738144
7 0.237508311867714
8 0.230111390352249
9 0.20174315571785
10 0.181914582848549
11 0.173216059803963
12 0.163795545697212
13 0.177881568670273
14 0.15107724070549
15 0.143447831273079
16 0.128239825367928
17 0.123446241021156
18 0.138343527913094
19 0.122009739279747
20 0.11266378313303
21 0.117821298539639
22 0.112564139068127
23 0.111942440271378
24 0.106780335307121
25 0.122175559401512
26 0.104393720626831
27 0.100206717848778
28 0.0950484350323677
29 0.0950669422745705
30 0.0912822708487511
31 0.0949487686157227
32 0.0905272290110588
33 0.0903654992580414
34 0.0908970534801483
35 0.0963043347001076
36 0.0999708622694016
37 0.105116099119186
38 0.115546010434628
39 0.0881784036755562
40 0.0875627994537354
41 0.087608315050602
42 0.0856340453028679
43 0.0852829813957214
44 0.0879020318388939
45 0.0785540342330933
46 0.0837889909744263
47 0.0912732928991318
48 0.096081517636776
49 0.0823747739195824
50 0.0811893790960312
51 0.0756397023797035
52 0.083668477833271
53 0.0794014558196068
54 0.0761493518948555
55 0.0776190385222435
56 0.0760229974985123
57 0.0775342956185341
58 0.0717881917953491
59 0.0866456180810928
60 0.080246813595295
61 0.0727540850639343
62 0.0846920534968376
63 0.073209285736084
64 0.0762777402997017
65 0.0738563686609268
66 0.0744405537843704
67 0.0729038193821907
68 0.0749403759837151
69 0.097957655787468
70 0.0880668386816978
71 0.0735488086938858
72 0.0738849639892578
73 0.0719012692570686
74 0.0676064342260361
75 0.0675804764032364
76 0.0647993534803391
77 0.066245898604393
78 0.0694892331957817
79 0.0739478766918182
80 0.0732904896140099
81 0.0708587914705276
82 0.0688145682215691
83 0.0631531849503517
84 0.064467154443264
85 0.0716394707560539
86 0.0648197308182716
87 0.0618649907410145
88 0.066649042069912
89 0.067033939063549
90 0.0711728185415268
91 0.0885690078139305
92 0.0761702880263329
93 0.0668672621250153
94 0.0706817060709
95 0.0656597912311554
96 0.0712591111660004
97 0.0681193917989731
98 0.0625659450888634
99 0.0598854385316372
100 0.0668763816356659
101 0.0676249265670776
102 0.0684297010302544
103 0.0695067346096039
104 0.0681540295481682
105 0.0649106651544571
106 0.0670194923877716
107 0.0614909939467907
108 0.0633795335888863
109 0.0687778368592262
110 0.0744949579238892
111 0.0680141299962997
112 0.0683108791708946
113 0.0619812570512295
114 0.0650999620556831
115 0.0611182600259781
116 0.0638155713677406
117 0.0638103410601616
118 0.0582005269825459
119 0.0629937797784805
120 0.0707551017403603
121 0.0645396709442139
122 0.0598633401095867
123 0.0620469078421593
124 0.0612957701086998
125 0.0625300779938698
126 0.0574979223310947
127 0.0700420811772346
128 0.0627750232815742
129 0.0561985336244106
130 0.0632045343518257
131 0.0667473822832108
132 0.0606741048395634
133 0.0629726499319077
134 0.0629926919937134
135 0.0604988411068916
136 0.0740029886364937
137 0.0839794054627419
138 0.0643352717161179
139 0.0575324818491936
140 0.0629472509026527
141 0.0636490657925606
142 0.0582785904407501
143 0.061954990029335
144 0.059744156897068
145 0.0637471750378609
146 0.0561295449733734
147 0.126853421330452
148 0.0604992471635342
149 0.057946939021349
150 0.0573049820959568
151 0.0588716454803944
152 0.0567386113107204
153 0.0593579262495041
154 0.0566021613776684
155 0.0567416995763779
156 0.0561171360313892
157 0.0622265450656414
158 0.0573424808681011
159 0.05408750846982
160 0.0581204034388065
161 0.0580939203500748
162 0.0580315180122852
163 0.0593780167400837
164 0.0585629343986511
165 0.0568298995494843
166 0.0627920627593994
167 0.0630361139774323
168 0.0555664449930191
169 0.0579487234354019
170 0.0585102997720242
171 0.0549028404057026
172 0.0562929473817348
173 0.0531077273190022
174 0.0583534464240074
175 0.0562650077044964
176 0.0603231005370617
177 0.0615253038704395
178 0.0557108148932457
179 0.0531161576509476
180 0.0578491389751434
181 0.057009432464838
182 0.0576318986713886
183 0.0569968670606613
184 0.0574024766683578
185 0.0570089854300022
186 0.0593397691845894
187 0.0553956218063831
188 0.0566733963787556
189 0.0566032715141773
190 0.0630143955349922
191 0.0549748316407204
};
\addlegendentry{Training Loss}
\addplot [line width=2pt, blue, dashed]
table {%
0 2.00816488265991
1 1.41219782829285
2 0.849621891975403
3 0.580694198608398
4 0.429149329662323
5 0.353573352098465
6 0.328401058912277
7 0.294911950826645
8 0.293363213539124
9 0.239846423268318
10 0.228754207491875
11 0.215639561414719
12 0.170335873961449
13 0.175098448991776
14 0.177863985300064
15 0.173584878444672
16 0.16017447412014
17 0.132818505167961
18 0.134418100118637
19 0.130954727530479
20 0.116741597652435
21 0.153949975967407
22 0.112711794674397
23 0.114618249237537
24 0.125530064105988
25 0.102457329630852
26 0.132476985454559
27 0.100840039551258
28 0.0995015949010849
29 0.0996487364172935
30 0.0934565588831902
31 0.0947203785181046
32 0.0886236578226089
33 0.116582661867142
34 0.109886601567268
35 0.0912807360291481
36 0.113108664751053
37 0.106487058103085
38 0.125551730394363
39 0.102875113487244
40 0.076774463057518
41 0.0873953327536583
42 0.0914807766675949
43 0.0919035524129868
44 0.0804400965571404
45 0.0802583545446396
46 0.0802309364080429
47 0.0886843651533127
48 0.0776375085115433
49 0.0811457335948944
50 0.078115813434124
51 0.0769429579377174
52 0.0731330141425133
53 0.0686452761292458
54 0.075811855494976
55 0.0745514780282974
56 0.0705928206443787
57 0.0665546059608459
58 0.0685306191444397
59 0.0693990513682365
60 0.0870403945446014
61 0.073952928185463
62 0.0671895816922188
63 0.0621306784451008
64 0.064313642680645
65 0.0680685117840767
66 0.0685467645525932
67 0.0648471787571907
68 0.0634088963270187
69 0.0790350884199142
70 0.0670035853981972
71 0.0632006749510765
72 0.0832783132791519
73 0.0702855288982391
74 0.0627079531550407
75 0.061639279127121
76 0.0623600445687771
77 0.0592284314334393
78 0.0858055949211121
79 0.0582492426037788
80 0.0580333806574345
81 0.0614533424377441
82 0.060555599629879
83 0.0591588690876961
84 0.0578888580203056
85 0.0638040602207184
86 0.0610727034509182
87 0.0555197782814503
88 0.0616348758339882
89 0.055053886026144
90 0.057916134595871
91 0.0638524144887924
92 0.0602666176855564
93 0.0587667748332024
94 0.057281132787466
95 0.0587302967905998
96 0.0569714531302452
97 0.057174377143383
98 0.0576536916196346
99 0.0586537346243858
100 0.0683775842189789
101 0.0572494640946388
102 0.0587028563022614
103 0.0652601793408394
104 0.060908131301403
105 0.0597879327833652
106 0.0570636168122292
107 0.0565977171063423
108 0.0583501979708672
109 0.0584105663001537
110 0.0604828149080276
111 0.05830167979002
112 0.0577057376503944
113 0.0565742738544941
114 0.0608632825314999
115 0.0564267039299011
116 0.0546353720128536
117 0.0547471791505814
118 0.0678886026144028
119 0.0556068383157253
120 0.0548176392912865
121 0.0564000681042671
122 0.0550211854279041
123 0.0592719353735447
124 0.0528233125805855
125 0.0598904266953468
126 0.0702000111341476
127 0.0563584417104721
128 0.0582901984453201
129 0.0534951463341713
130 0.114254415035248
131 0.0531537644565105
132 0.0566905401647091
133 0.0639152303338051
134 0.0529140569269657
135 0.0551792643964291
136 0.134003043174744
137 0.0694111064076424
138 0.0559732504189014
139 0.054705273360014
140 0.0731992274522781
141 0.053338672965765
142 0.0554924607276917
143 0.0551909394562244
144 0.0528316162526608
145 0.0548948980867863
146 0.0582743547856808
147 0.0546383038163185
148 0.0557272136211395
149 0.055961512029171
150 0.0494087636470795
151 0.0547849014401436
152 0.0548192113637924
153 0.0593798793852329
154 0.0564047135412693
155 0.0566342957317829
156 0.0601980425417423
157 0.0520054139196873
158 0.0541595481336117
159 0.0553053729236126
160 0.0553874112665653
161 0.0541112013161182
162 0.0529037825763226
163 0.0531395524740219
164 0.054907314479351
165 0.0547817647457123
166 0.0552332401275635
167 0.0553476624190807
168 0.0560474321246147
169 0.0536957643926144
170 0.0538466237485409
171 0.0581373758614063
172 0.0536079332232475
173 0.0523171685636044
174 0.0530385673046112
175 0.0531677268445492
176 0.0521033704280853
177 0.0523441322147846
178 0.05159867182374
179 0.0529480129480362
180 0.0537430867552757
181 0.0524222701787949
182 0.0534558519721031
183 0.05147710070014
184 0.0527499690651894
185 0.0531826503574848
186 0.0521434582769871
187 0.0575625970959663
188 0.0514901503920555
189 0.0519134476780891
190 0.0544096417725086
191 0.0530923008918762
};
\addlegendentry{Validation Loss}
\end{axis}

\begin{axis}[
width=3.7in,
height=2.1in,
axis y line=right,
legend cell align={left},
legend columns=1,
legend style={
  fill opacity=0.8,
  draw opacity=1,
  text opacity=1,
  at={(0.7,0.62)},
  anchor=south,
  draw=lightgray204
},
tick align=outside,
x grid style={darkgray176},
xmin=-9.55, xmax=200.55,
xtick pos=left,
xtick style={color=black},
y grid style={darkgray176},
ylabel={\textcolor{red} {Accuracy}},
ymin=0.0650298453867435, ymax=1.03372219130397,
ytick pos=right,
ytick style={color=red},
yticklabel style={anchor=west,color=red}
]
\addplot [line width=2pt, red]
table {%
0 0.288454592227936
1 0.448425620794296
2 0.740499436855316
3 0.868621051311493
4 0.913861751556396
5 0.923633754253387
6 0.913137912750244
7 0.948606610298157
8 0.946073114871979
9 0.95439738035202
10 0.959826290607452
11 0.962359726428986
12 0.967064797878265
13 0.956568956375122
14 0.966702878475189
15 0.969236314296722
16 0.975389063358307
17 0.973579466342926
18 0.967064797878265
19 0.976112902164459
20 0.976836800575256
21 0.975750982761383
22 0.974665224552155
23 0.977560639381409
24 0.975389063358307
25 0.97321754693985
26 0.978284478187561
27 0.977922558784485
28 0.980456054210663
29 0.980456054210663
30 0.981179893016815
31 0.979008316993713
32 0.980456054210663
33 0.980456054210663
34 0.981179893016815
35 0.979732155799866
36 0.977198719978333
37 0.976474821567535
38 0.973941385746002
39 0.983351409435272
40 0.98262757062912
41 0.981903731822968
42 0.981541812419891
43 0.98262757062912
44 0.980817973613739
45 0.982989490032196
46 0.982989490032196
47 0.981541812419891
48 0.976836800575256
49 0.981179893016815
50 0.983351409435272
51 0.983351409435272
52 0.981903731822968
53 0.984075307846069
54 0.982989490032196
55 0.984075307846069
56 0.983713328838348
57 0.981179893016815
58 0.984075307846069
59 0.977922558784485
60 0.980817973613739
61 0.985161066055298
62 0.981541812419891
63 0.984437227249146
64 0.98262757062912
65 0.983351409435272
66 0.983713328838348
67 0.982989490032196
68 0.982265651226044
69 0.973579466342926
70 0.978646397590637
71 0.984075307846069
72 0.982265651226044
73 0.983713328838348
74 0.98588490486145
75 0.983713328838348
76 0.984799146652222
77 0.986608743667603
78 0.984437227249146
79 0.98262757062912
80 0.982989490032196
81 0.983713328838348
82 0.985161066055298
83 0.985522985458374
84 0.984075307846069
85 0.983713328838348
86 0.985161066055298
87 0.983713328838348
88 0.985161066055298
89 0.985522985458374
90 0.984075307846069
91 0.976112902164459
92 0.982989490032196
93 0.982989490032196
94 0.984075307846069
95 0.983713328838348
96 0.981903731822968
97 0.984075307846069
98 0.986246824264526
99 0.984799146652222
100 0.984437227249146
101 0.983351409435272
102 0.982989490032196
103 0.983351409435272
104 0.983713328838348
105 0.985161066055298
106 0.984075307846069
107 0.985161066055298
108 0.983713328838348
109 0.983351409435272
110 0.982989490032196
111 0.984437227249146
112 0.983713328838348
113 0.984799146652222
114 0.983351409435272
115 0.986608743667603
116 0.984437227249146
117 0.984075307846069
118 0.98588490486145
119 0.984799146652222
120 0.983713328838348
121 0.982989490032196
122 0.98588490486145
123 0.984437227249146
124 0.985161066055298
125 0.983713328838348
126 0.986246824264526
127 0.982989490032196
128 0.984799146652222
129 0.983351409435272
130 0.984437227249146
131 0.983351409435272
132 0.985522985458374
133 0.983351409435272
134 0.984075307846069
135 0.985161066055298
136 0.981179893016815
137 0.977560639381409
138 0.982989490032196
139 0.984799146652222
140 0.984799146652222
141 0.983713328838348
142 0.985522985458374
143 0.984437227249146
144 0.984799146652222
145 0.983713328838348
146 0.985522985458374
147 0.960550129413605
148 0.982265651226044
149 0.984799146652222
150 0.985161066055298
151 0.986246824264526
152 0.98588490486145
153 0.985161066055298
154 0.985522985458374
155 0.98588490486145
156 0.986970663070679
157 0.984437227249146
158 0.985161066055298
159 0.986246824264526
160 0.984075307846069
161 0.984799146652222
162 0.986246824264526
163 0.985161066055298
164 0.984075307846069
165 0.986246824264526
166 0.984799146652222
167 0.983351409435272
168 0.98588490486145
169 0.984799146652222
170 0.985161066055298
171 0.985161066055298
172 0.986246824264526
173 0.987332582473755
174 0.985161066055298
175 0.985161066055298
176 0.984075307846069
177 0.985522985458374
178 0.985522985458374
179 0.987332582473755
180 0.985161066055298
181 0.985522985458374
182 0.985522985458374
183 0.986608743667603
184 0.982989490032196
185 0.985522985458374
186 0.984437227249146
187 0.985522985458374
188 0.984075307846069
189 0.986246824264526
190 0.984075307846069
191 0.985161066055298
};
\addlegendentry{Training Accuracy}
\addplot [line width=2pt, red, dashed]
table {%
0 0.109061315655708
1 0.417254477739334
2 0.724905073642731
3 0.903418362140656
4 0.940314710140228
5 0.943027675151825
6 0.947368443012238
7 0.943570256233215
8 0.942485094070435
9 0.955507338047028
10 0.954964756965637
11 0.954964756965637
12 0.974498093128204
13 0.972327709197998
14 0.968529582023621
15 0.965274035930634
16 0.970157325267792
17 0.981009244918823
18 0.97558331489563
19 0.976668477058411
20 0.981551826000214
21 0.968529582023621
22 0.982094407081604
23 0.981551826000214
24 0.977211058139801
25 0.981551826000214
26 0.972870349884033
27 0.981551826000214
28 0.981009244918823
29 0.982094407081604
30 0.981551826000214
31 0.982094407081604
32 0.983179569244385
33 0.977753639221191
34 0.977211058139801
35 0.98426479101181
36 0.977211058139801
37 0.980466604232788
38 0.97558331489563
39 0.978296279907227
40 0.986435174942017
41 0.98372220993042
42 0.981009244918823
43 0.982094407081604
44 0.984807372093201
45 0.986435174942017
46 0.984807372093201
47 0.982094407081604
48 0.987520337104797
49 0.981551826000214
50 0.984807372093201
51 0.985349953174591
52 0.986435174942017
53 0.986435174942017
54 0.984807372093201
55 0.985349953174591
56 0.987520337104797
57 0.985349953174591
58 0.987520337104797
59 0.985892593860626
60 0.982636988162994
61 0.982636988162994
62 0.98426479101181
63 0.987520337104797
64 0.986977756023407
65 0.985892593860626
66 0.985349953174591
67 0.986977756023407
68 0.984807372093201
69 0.985349953174591
70 0.98426479101181
71 0.984807372093201
72 0.977753639221191
73 0.983179569244385
74 0.987520337104797
75 0.987520337104797
76 0.985349953174591
77 0.989690721035004
78 0.981551826000214
79 0.987520337104797
80 0.987520337104797
81 0.984807372093201
82 0.985349953174591
83 0.988062918186188
84 0.986977756023407
85 0.984807372093201
86 0.984807372093201
87 0.986435174942017
88 0.988062918186188
89 0.986977756023407
90 0.987520337104797
91 0.984807372093201
92 0.986977756023407
93 0.985892593860626
94 0.985892593860626
95 0.984807372093201
96 0.985349953174591
97 0.986435174942017
98 0.987520337104797
99 0.98426479101181
100 0.98372220993042
101 0.986435174942017
102 0.986977756023407
103 0.98426479101181
104 0.985349953174591
105 0.986977756023407
106 0.985892593860626
107 0.986435174942017
108 0.988062918186188
109 0.984807372093201
110 0.985892593860626
111 0.983179569244385
112 0.986435174942017
113 0.985349953174591
114 0.986435174942017
115 0.985892593860626
116 0.986435174942017
117 0.987520337104797
118 0.985892593860626
119 0.985892593860626
120 0.988062918186188
121 0.986435174942017
122 0.986435174942017
123 0.985892593860626
124 0.985892593860626
125 0.985349953174591
126 0.985349953174591
127 0.985892593860626
128 0.985349953174591
129 0.985892593860626
130 0.967444360256195
131 0.986435174942017
132 0.986435174942017
133 0.981009244918823
134 0.986435174942017
135 0.986435174942017
136 0.96798700094223
137 0.985892593860626
138 0.986435174942017
139 0.984807372093201
140 0.977753639221191
141 0.986435174942017
142 0.985892593860626
143 0.986435174942017
144 0.985892593860626
145 0.985892593860626
146 0.98372220993042
147 0.98372220993042
148 0.986435174942017
149 0.986435174942017
150 0.988062918186188
151 0.987520337104797
152 0.985892593860626
153 0.985349953174591
154 0.987520337104797
155 0.986435174942017
156 0.985349953174591
157 0.986435174942017
158 0.986435174942017
159 0.985892593860626
160 0.985892593860626
161 0.984807372093201
162 0.985892593860626
163 0.986435174942017
164 0.986435174942017
165 0.986435174942017
166 0.985892593860626
167 0.98426479101181
168 0.98426479101181
169 0.984807372093201
170 0.985892593860626
171 0.98426479101181
172 0.986435174942017
173 0.986435174942017
174 0.984807372093201
175 0.986435174942017
176 0.985892593860626
177 0.986435174942017
178 0.985892593860626
179 0.985892593860626
180 0.984807372093201
181 0.98426479101181
182 0.984807372093201
183 0.986435174942017
184 0.986435174942017
185 0.986435174942017
186 0.985892593860626
187 0.985349953174591
188 0.986435174942017
189 0.985892593860626
190 0.984807372093201
191 0.985892593860626
};
\addlegendentry{Validation Accuracy}
\end{axis}

\end{tikzpicture}

%% file: figures/msg1_distvscent_average_accuracy.tikz
% This file was created with tikzplotlib v0.10.1.
\begin{tikzpicture}

\definecolor{darkgray176}{RGB}{176,176,176}
\definecolor{darkorange25512714}{RGB}{255,127,14}
\definecolor{lightgray204}{RGB}{204,204,204}
\definecolor{steelblue31119180}{RGB}{31,119,180}

\begin{axis}[
width=3.7in,
height=2.1in,
legend cell align={left},
legend style={
  fill opacity=0.8,
  draw opacity=1,
  text opacity=1,
  at={(0.97,0.03)},
  anchor=south east,
  draw=lightgray204
},
tick align=outside,
tick pos=left,
x grid style={darkgray176,dashed},
xlabel={Lookback ($\ell$)},
xmin=0, xmax=10,
xmajorgrids,
xtick style={color=black},
y grid style={darkgray176,dashed},
ylabel={Average Accuracy ($\bar A$)},
ymin=0, ymax=0.8,
ymajorgrids,
ytick style={color=black}
]
\addplot [line width=2pt, mark=square, mark size=3, mark repeat=1, mark phase=0, mark options={solid}, steelblue31119180]
table {%
0 0.619399914211371
1 0.685282241170373
2 0.705034062644846
3 0.712239466546515
4 0.737613028759184
5 0.735353709154844
6 0.75484301533531
7 0.765399510706129
8 0.769293304266343
9 0.787235244860973
10 0.78202804480415
};
\addlegendentry{Distributed (no coords)}
\addplot [line width=2pt, mark=triangle, mark size=3, mark repeat=1, mark phase=0, mark options={solid}, darkorange25512714]
table {%
0 0.136005128965301
1 0.322653725989401
2 0.459000737739397
3 0.58354314272981
4 0.582562931104917
5 0.609731033885993
6 0.578363947428412
7 0.603433836485678
8 0.617353435800808
9 0.602409579937685
10 0.607358392666355
};
\addlegendentry{Centralized}
\addplot [line width=2pt ,mark=*, gray]
table {%
0 0.619399914211371
1 0.619399914211371
2 0.619399914211371
3 0.619399914211371
4 0.619399914211371
5 0.619399914211371
6 0.619399914211371
7 0.619399914211371
8 0.619399914211371
9 0.619399914211371
10 0.619399914211371
};
\addlegendentry{Legacy \hl{(no deep learning)}}
\end{axis}

\end{tikzpicture}

%% file: figures/msg2_acc_.tikz
\begin{tikzpicture}
% Generated from XXXX  using varepsilon = 0.03
\definecolor{darkgray176}{RGB}{176,176,176}

\begin{axis}[
width=3.7in,
height=2.1in,
tick align=outside,
tick pos=left,
x grid style={darkgray176,dashed},
xlabel={Lookback ($\ell$)},
xmin=0, xmax=10,
xmajorgrids,
xtick style={color=black},
y grid style={darkgray176,dashed},
ylabel={$\boldsymbol\alpha(\varepsilon=0.3)$},
ytick={0,0.25,...,1},
ymin=-0.1, ymax=1.1,
ymajorgrids,
ytick style={color=black}
]
\addplot[blue, mark=*, mark size=3, mark repeat=1, line width=2pt] table {%
0 0 
1 0
2 1
3 1
4 1
5 1
6 1
7 1 
8 0 
9 0 
10 0
};

% \addplot+[const plot, mark=*, mark size=3, mark repeat=1, line width=2pt] coordinates
% {
% (0,0)
% (1,0)
% (2,1)
% (3,1)
% (4,1)
% (5,1)
% (6,1)
% (7,1)
% (8,0)
% (9,0)
% (10,0)
% };
\end{axis}

\end{tikzpicture}

%% file: figures/coherence.tikz
% This file was created with tikzplotlib v0.10.1.
\begin{tikzpicture}

\definecolor{darkgray176}{RGB}{176,176,176}

\begin{axis}[
width=3.7in,
height=2.1in,
legend style={fill opacity=0.8, draw opacity=1, text opacity=1, draw=lightgray204},
tick align=outside,
tick pos=left,
x grid style={darkgray176,dashed},
xlabel={Coherence Time [Frames]},
xmajorgrids,
xmin=0, xmax=10,
xtick={0,1,...,10},
xtick style={color=black},
y grid style={darkgray176,dashed},
ylabel={CDF},
ymajorgrids,
ymin=0, ymax=1.05,
ytick={0,0.2,...,1},
ytick style={color=black}
]
\addplot [line width=2pt, black]
table {%
0.488035045952523 0.00160232590937974
0.439994901066406 0.000226860203378496
0.536075190838641 0.0064649570273133
0.584115335724758 0.0175826715459864
0.632155480610875 0.0351599789098238
0.680195625496992 0.0545868115770673
0.728235770383109 0.0717606877417933
0.776275915269226 0.085201540146098
0.824316060155344 0.0962470617133511
0.872356205041461 0.106607756026268
0.920396349927578 0.117381715872235
0.968436494813695 0.12853675615836
1.01647663969981 0.13975706066063
1.06451678458593 0.150800123644399
1.11255692947205 0.161732326863757
1.16059707435816 0.172563504652646
1.20863721924428 0.183217440923033
1.2566773641304 0.193718498002177
1.30471750901652 0.204050583343633
1.35275765390263 0.214143739071876
1.40079779878875 0.223931806940402
1.44883794367487 0.233354886915218
1.49687808856098 0.242475337613802
1.5449182334471 0.251305675461169
1.59295837833322 0.259954301638244
1.64099852321934 0.268422557190563
1.68903866810545 0.27691159894871
1.73707881299157 0.285588610589656
1.78511895787769 0.29442990030891
1.8331591027638 0.303265825846015
1.88119924764992 0.312338892935618
1.92923939253604 0.321559028019135
1.97727953742216 0.330780727655779
2.02531968230827 0.340100994139407
2.07335982719439 0.349526532697708
2.12139997208051 0.359123725084771
2.16944011696662 0.368832224251423
2.21748026185274 0.378589001057414
2.26552040673886 0.388234918098997
2.31356055162498 0.397864966101721
2.36160069651109 0.407436678623578
2.40964084139721 0.416956090369483
2.45768098628333 0.426381405420193
2.50572113116945 0.43571240026812
2.55376127605556 0.444964273429351
2.60180142094168 0.454168092458831
2.6498415658278 0.463340396918188
2.69788171071391 0.472409887886358
2.74592185560003 0.481402268736138
2.79396200048615 0.490414094746208
2.84200214537227 0.499427708816994
2.89004229025838 0.508415172499805
2.9380824351445 0.517456054496514
2.98612258003062 0.526519510759767
3.03416272491673 0.535517032284107
3.08220286980285 0.544501590368253
3.13024301468897 0.553554988789977
3.17828315957509 0.562569273383531
3.2263233044612 0.571621777774897
3.27436344934732 0.580705796736387
3.32240359423344 0.589825353404613
3.37044373911955 0.599025372805072
3.41848388400567 0.608118555577732
3.46652402889179 0.617181341318216
3.51456417377791 0.626209706889912
3.56260431866402 0.635388493069365
3.61064446355014 0.644553645285856
3.65868460843626 0.653593186237028
3.70672475332237 0.662631386142663
3.75476489820849 0.671662210297844
3.80280504309461 0.680619053440888
3.85084518798073 0.689584389872334
3.89888533286684 0.698427020637126
3.94692547775296 0.707236125263487
3.99496562263908 0.715994940682203
4.0430057675252 0.724730958326786
4.09104591241131 0.733402605785583
4.13908605729743 0.742041621136308
4.18712620218355 0.750604643906591
4.23516634706966 0.759085192376335
4.28320649195578 0.767450410929878
4.3312466368419 0.775654480511367
4.37928678172802 0.78363034884394
4.42732692661413 0.791408412959774
4.47536707150025 0.798972580312423
4.52340721638637 0.806361294207286
4.57144736127248 0.813573660614006
4.6194875061586 0.820543297778491
4.66752765104472 0.82724606688107
4.71556779593084 0.833725998916882
4.76360794081695 0.839894584880471
4.81164808570307 0.845842568853189
4.85968823058919 0.851630744899389
4.9077283754753 0.85725486637487
4.95576852036142 0.862579934695552
5.00380866524754 0.867717703656204
5.05184881013366 0.872647163543408
5.09988895501977 0.877366749804039
5.14792909990589 0.881930104259585
5.19596924479201 0.886226143638046
5.24400938967813 0.890258444060855
5.29204953456424 0.894159098513428
5.34008967945036 0.897919390199773
5.38812982433648 0.90144768100818
5.43616996922259 0.904929035222784
5.48421011410871 0.908274496822951
5.53225025899483 0.911454339299272
5.58029040388095 0.914618759751915
5.62833054876706 0.917603927117751
5.67637069365318 0.920484269424094
5.7244108385393 0.923285937058921
5.77245098342541 0.926011612113306
5.82049112831153 0.928558034080884
5.86853127319765 0.931049026166257
5.91657141808377 0.933479671202455
5.96461156296988 0.935831418059547
6.012651707856 0.938145839149187
6.06069185274212 0.940394996022682
6.10873199762823 0.942560784565281
6.15677214251435 0.944633146935454
6.20481228740047 0.946640468597072
6.25285243228659 0.948654718993965
6.3008925771727 0.95054335812554
6.34893272205882 0.952414116649952
6.39697286694494 0.954151664651001
6.44501301183105 0.955826630526979
6.49305315671717 0.957439461293067
6.54109330160329 0.958984345751936
6.58913344648941 0.960478046972802
6.63717359137552 0.961945374298102
6.68521373626164 0.963388562803733
6.73325388114776 0.964764252017323
6.78129402603388 0.966132565480459
6.82933417091999 0.967512054323072
6.87737431580611 0.968894672271939
6.92541446069223 0.97015480806174
6.97345460557834 0.971321741186704
7.02149475046446 0.972444419808941
7.06953489535058 0.97353714841418
7.1175750402367 0.974548296749238
7.16561518512281 0.975489040193593
7.21365533000893 0.976378600399944
7.26169547489505 0.977264584484863
7.30973561978116 0.978111231234023
7.35777576466728 0.978952066785855
7.4058159095534 0.979743060145221
7.45385605443952 0.980494716168829
7.50189619932563 0.981237878904035
7.54993634421175 0.981968972229405
7.59797648909787 0.982663410310092
7.64601663398398 0.983322981206811
7.6940567788701 0.983952602086532
7.74209692375622 0.984560766237658
7.79013706864234 0.985155296425823
7.83817721352845 0.985727475855033
7.88621735841457 0.986276410494933
7.93425750330069 0.986818863414735
7.9822976481868 0.987336506992099
8.03033779307292 0.987842528174807
8.07837793795904 0.988332456811069
8.12641808284516 0.988810092529906
8.17445822773127 0.989278564437573
8.22249837261739 0.989726250139412
8.27053851750351 0.990160972401059
8.31857866238962 0.990591001003325
8.36661880727574 0.991013430347547
8.41465895216186 0.991412838410047
8.46269909704798 0.991805541244861
8.51073924193409 0.992189080268504
8.55877938682021 0.992562337943028
8.60681953170633 0.992934254572015
8.65485967659244 0.993287620071071
8.70289982147856 0.993631821758955
8.75093996636468 0.993965071574953
8.7989801112508 0.994279546753429
8.84702025613691 0.994589775287704
8.89506040102303 0.994888381427323
8.94310054590915 0.995177153233003
8.99114069079526 0.995463019440019
9.03918083568138 0.995738604297916
9.0872209805675 0.996009048481254
9.13526112545362 0.99627099937619
9.18330127033973 0.996524233475134
9.23134141522585 0.996775008990592
9.27938156011197 0.99701662069488
9.32742170499809 0.997247057019691
9.3754618498842 0.997474140730659
9.42350199477032 0.99769988339609
9.47154213965644 0.997914227174455
9.51958228454255 0.998130135505946
9.56762242942867 0.998338668086982
9.61566257431479 0.998547200668019
9.66370271920091 0.998744781377168
9.71174286408702 0.998938562457296
9.75978300897314 0.999127426370453
9.80782315385926 0.999310926101462
9.85586329874537 0.999490179188269
9.90390344363149 0.999664291600517
9.95194358851761 0.999829910724363
9.99998373340373 0.999999999999999
};
\end{axis}

\end{tikzpicture}

%% file: figures/msg3_dist_faris_excel.tikz
\begin{tikzpicture}

\definecolor{crimson2143940}{RGB}{214,39,40}
\definecolor{darkgray176}{RGB}{176,176,176}
\definecolor{darkorange25512714}{RGB}{255,127,14}
\definecolor{darkturquoise23190207}{RGB}{23,190,207}
\definecolor{forestgreen4416044}{RGB}{44,160,44}
\definecolor{goldenrod18818934}{RGB}{188,189,34}
\definecolor{gray127}{RGB}{127,127,127}
\definecolor{lightgray204}{RGB}{204,204,204}
\definecolor{mediumpurple148103189}{RGB}{148,103,189}
\definecolor{orchid227119194}{RGB}{227,119,194}
\definecolor{sienna1408675}{RGB}{140,86,75}
\definecolor{steelblue31119180}{RGB}{31,119,180}

\begin{axis}[
width=3.7in,
height=2.1in,
legend cell align={left},
legend style={
  fill opacity=0.8,
  draw opacity=1,
  text opacity=1,
  at={(0.67,0.21)},
  anchor=west,
  draw=lightgray204
},
tick align=outside,
tick pos=left,
x grid style={darkgray176,dashed},
xlabel={Lookback ($\ell$)},
xmin=0, xmax=10,
xmajorgrids,
xtick style={color=black},
xtick={-2,0,2,4,6,8,10,12},
y grid style={darkgray176,dashed},
ylabel={Accuracy},
ymin=0.2, ymax=1,
ymajorgrids,
ytick style={color=black}
]
\addlegendentry{Minimum}
\addplot [line width=2pt, mark=square, mark size=3, mark repeat=2, mark phase=0, mark options={solid}, crimson2143940]
table {%
0 0.320038270645219
1 0.47413741553549
2 0.586990037950664
3 0.517611480075901
4 0.564141996667948
5 0.5853889943074
6 0.606024667931689
7 0.604660815939279
8 0.58260955388867
9 0.601755218216319
10 0.595706831119545
};

\addlegendentry{Maximum}
\addplot [line width=2pt, mark=triangle, mark size=3, mark repeat=2, mark phase=0, mark options={solid}, mediumpurple148103189]
table {%
0 0.911965645129807
1 0.931290259613508
2 0.924653523326176
3 0.932461448370096
4 0.923091938317392
5 0.93011907085692
6 0.88120283558198
7 0.941050165918407
8 0.923677532695686
9 0.940269373414015
10 0.939488580909623
};

\addlegendentry{Average}
\addplot [line width=2pt, mark=*, mark size=3, mark repeat=2, mark phase=0, mark options={solid}, steelblue31119180]
table {%
 0 0.617391057614355
1 0.68293877418445
2 0.726169049828187
3 0.728120313780124
4 0.741192458242379
5 0.750718850451274
6 0.7599466484376
7 0.763720918362404
8 0.766332019827494
9 0.787444986681781
10 0.791693313615281
};

\end{axis}
\end{tikzpicture}